\documentclass[journal,12pt,onecolumn,draftclsnofoot,a4paper,]{IEEEtran}

\usepackage{cite}
\usepackage[T1]{fontenc}
\usepackage[latin1]{inputenc}
\usepackage{graphicx}
\usepackage{theorem}
\usepackage{amsmath}
\usepackage{amsfonts}
\usepackage{amssymb }

\usepackage{algorithm}
\usepackage{algpseudocode}

\ifCLASSOPTIONcompsoc
\usepackage[caption=false,font=normalsize,labelfon
t=sf,textfont=sf]{subfig}
\else
\usepackage[caption=false,font=footnotesize]{subfi
g}
\fi
\usepackage[acronym,shortcuts]{glossaries}
\usepackage{color}
\usepackage{url}
\usepackage{enumerate}
\usepackage{comment}
\usepackage{cleveref}
\usepackage{multirow}
\usepackage{color}

\newacronym{BSC}{BSC}{binary symmetric channel}
\newacronym{DFR}{DFR}{decryption failure rate}
\newacronym{ISD}{ISD}{information set decoding}
\newacronym{QC}{QC}{quasi-cyclic}
\newacronym{QC-LDPC}{QC-LDPC}{quasi-cyclic low-density parity-check}
\newacronym{QC-MDPC}{QC-MDPC}{quasi-cyclic moderate-density parity-check}
\newacronym{QC-LDGM}{QC-LDGM}{quasi-cyclic low-density generator matrix}
\newacronym{LDPC}{LDPC}{low-density parity-check}
\newacronym{MDPC}{MDPC}{moderate-density parity-check}
\newacronym{LDPCC}{LDPCC}{low-density parity-check convolutional}
\newacronym{NP}{NP}{non-polynomial}
\newacronym{QD}{QD}{quasi-dyadic}
\newacronym{GRS}{GRS}{generalized Reed-Solomon}
\newacronym{AC-LDPC}{AC-LDPC}{array convolutional low-density parity-check}
\newacronym{PDC-LDPC}{PDC-LDPC}{progressive differences convolutional low-density parity-check}
\newacronym{SC-LDPC}{SC-LDPC}{spatially coupled low-density parity-check}
\newacronym{DLP}{DLP}{Discrete Logarithm Problem}
\newacronym{IFP}{IFP}{Integer Factorization Problem}
\newacronym{SC-LDPC-CCs}{SC-LDPC-CCs}{spatially coupled low-density parity-check convolutional codes}
\newacronym{SC-LDPC-CC}{SC-LDPC-CC}{spatially coupled low-density parity-check convolutional code}
\newacronym{AWGN}{AWGN}{additive white Gaussian noise}
\newacronym{BF}{BF}{bit flipping}
\newacronym{BER}{BER}{bit error rate}
\newacronym{CER}{CER}{codeword error rate}
\newacronym{FER}{FER}{frame error rate}
\newacronym{TUB}{TUB}{truncated union bound}
\newacronym{BPSK}{BPSK}{binary phase shift keying}
\newacronym{SPA-LLR}{SPA-LLR}{sum-product algorithm with log-likelihood ratios}
\newacronym{RTI}{RTI}{regular time-invariant}
\newacronym{RTI-LDPCC}{RTI-LDPCC}{regular time-invariant low-density parity-check convolutional}
\newacronym{CPA}{CPA}{chosen-plaintext attack}
\newacronym{CCA2}{CCA2}{adaptive chosen-ciphertext attack}
\newacronym{SL}{SL}{security level}
\newacronym{BEC}{BEC}{binary erasure channel}
\newacronym{CH-GLDPC}{CH-GLDPC}{check-hybrid generalized low-density parity-check}
\newacronym{IND-CCA}{IND-CCA}{indistinguishability under adaptive chosen ciphertext attack}

{\theorembodyfont{\rmfamily}
   }
{\theorembodyfont{\rmfamily}
   }
{\theorembodyfont{\rmfamily}
   \newtheorem{Lem}{{\textbf Lemma}}}
{\theorembodyfont{\rmfamily}
   \newtheorem{Cor}{{\textbf Corollary}}}
{\theorembodyfont{\rmfamily}
   }
{\theorembodyfont{\rmfamily}
   }
{\theorembodyfont{\rmfamily}
   }

\newtheorem{theorem}{Theorem}

\def\HH{\mathbf{H}}

\def\0{\bar{0}}

\newcommand{\bH}{\mathbf{H}}
\newcommand{\bh}{\mathbf{h}}
\newcommand{\bs}{\mathbf{s}}
\newcommand{\be}{\mathbf{e}}
\newcommand{\bGamma}{\boldsymbol{\Gamma}}
\newcommand{\ba}{\mathbf{a}}
\newcommand{\bgamma}[1]{\boldsymbol{\gamma}^{(#1)}}

\newcommand{\Bt}{\mathcal{B}_t}
\newcommand{\st}{\text{\hspace{1.5mm}s.t.\hspace{1.5mm}}}

\newcommand{\errset}[1]{\mathcal{E}^{#1}_{i,t,b_i}}
\newcommand{\card}[1]{\left|#1\right|}
\newcommand{\btgamma}[1]{\tilde{\boldsymbol{\gamma}}^{(#1)}}

\newcommand{\Nsub}[3]{\mathcal{N}^{#1}_{#2,#3}}

\newcommand{\bmaj}{\left\lceil\frac{v}{2}\right\rceil}

\newcommand{\weight}[1]{\mathrm{wt}\left( #1\right)}
\makeatletter
\newcommand{\tpmod}[1]{{\@displayfalse\pmod{#1}}}
\makeatother

\definecolor{ps}{rgb}{1,0,0}
\definecolor{cc}{rgb}{0,0,1}
\newtheorem{defn}{Definition}
\newtheorem{prob}{Problem}

\begin{document}

\title{Analysis of the error correction capability of {LDPC} and {MDPC} codes under parallel bit-flipping decoding and application to cryptography\thanks{The  material  in  this  paper  has  been  presented  in  part  at  the  2019  IEEE  International  Conference on Communications, Shanghai (China) \cite{Santini2019}.}}

\author{\IEEEauthorblockN{Paolo Santini, Massimo Battaglioni, Marco Baldi and Franco Chiaraluce}\\
\IEEEauthorblockA{Dipartimento di Ingegneria dell'Informazione\\
Universit\`a Politecnica delle Marche\\
Ancona, Italy, 60131\\
Email:  p.santini@pm.univpm.it, \{m.battaglioni, m.baldi, f.chiaraluce\}@staff.univpm.it}}


\maketitle
\begin{abstract}
Iterative decoders used for decoding low-density parity-check (LDPC) and moderate-density parity-check (MDPC) codes are not characterized by a deterministic decoding radius and their error rate performance is usually assessed through intensive Monte Carlo simulations. 
However, several applications, like code-based cryptography, need guaranteed low values of the error rate, which are infeasible to assess through simulations, thus requiring the development of theoretical models for the error rate of these codes under iterative decoding. Some models of this type already exist, but become computationally intractable for parameters of practical interest. 
Other approaches approximate the code ensemble behaviour through some assumptions, which may not hold true for a specific code.
We propose a theoretical analysis of the error correction capability of LDPC and MDPC codes that allows deriving tight bounds on the error rate at the output of parallel bit-flipping decoders. Special attention is devoted to the case of codes with small girth; moreover, single-iteration decoding is investigated through a rigorous approach, which  does not require any assumption and hence results in a guaranteed error correction capability for any single code.
We show an example of application of the new bound to the context of code-based cryptography, where guaranteed error rates are needed to achieve some strong security levels.

\end{abstract}

\begin{IEEEkeywords}
Bit flipping decoder, code-base cryptography, error correction capability, LDPC codes, MDPC codes.
\end{IEEEkeywords}

\section{Introduction}

Contrary to bounded distance decoders, iterative decoders commonly used for \ac{LDPC} and \ac{MDPC} codes are not characterized by a deterministic decoding radius. 
This implies the existence of a residual error rate that is difficult to model theoretically, and is hence usually assessed through Monte Carlo simulations.
Nevertheless, there are applications in which extremely low error rates are required.
One of these cases is in the area of code-based cryptography, where error rates as low as $2^{-80}$ or less are required to avoid some types of attacks \cite{Fabsic2017, Paiva2018, Eaton2018, Santini2019a}.
Obviously, such low values of the error rate are infeasible to assess through numerical simulations.

Therefore, an important research challenge is represented by the development of analytical tools able to foresee the number of errors that an iterative decoder can correct. A vast body of literature exists on this subject \cite{Zyablov1975, Burshtein2008, Chilappagari2008,Chilappagari2009b,Chen2011}, which permits to determine lower and upper bounds on the guaranteed error correction capability of the code. Many of these approaches use expander graph based arguments \cite{Chilappagari2008,Chilappagari2009b}, whose application, however, is known to be NP-hard \cite{Alon1998} and can be used for a limited number of cases and under specific constraints. Moreover, the bounds these methods provide are often loose, particularly in case of small girths.

To overcome these limitations, recently, in \cite{Tillich2018} and \cite{Santini2019}, a new approach has been proposed to evaluate the guaranteed error correction capability of \ac{LDPC} and \ac{MDPC} codes. In \cite{Tillich2018}, in particular, a majority-logic decoder is considered and it is shown that its error correction capability depends on the maximum number of superimpositions between any two columns of the code parity-check matrix. This allows deriving conditions under which a single iteration of this decoder corrects all errors up to a given weight. These results are extended in \cite{Santini2019}, where a more general decoder is considered and tighter bounds are derived.

The latter results, however, are obtained under some assumptions. As a first contribution, this paper improves the analysis in \cite{Santini2019}, by providing tighter bounds.  For such a purpose, we focus attention on Gallager's \ac{BF} decoder \cite{Gallager1963}, because of its high computational efficiency, due to a relatively low algorithmic complexity.

Low-complexity iterative decoders are important in many applications where high throughputs have to be achieved. Starting from its basic principle, several variants of Gallager's \ac{BF} algorithm have been proposed. Among them, in this paper we focus on the so-called parallel \ac{BF}. Roughly speaking, the parallel \ac{BF} algorithm operates as follows. At each iteration, all parity checks are computed: all bits involved in a number of unsatisfied parity-check equations overcoming some suitably chosen threshold are flipped, and the syndrome is accordingly updated. The procedure is iterated, until a null syndrome is obtained or a maximum number of iterations is reached. Following a more general approach than \cite{Sipser1996}, where parallel \ac{BF} is introduced, we consider a threshold that is not fixed, but rather depends on some features of the code under investigation.

In principle, other families of iterative decoding algorithms could achieve better error correction performance than \ac{BF} decoding.
However, we focus on channel models without soft information, where decoding algorithms working with discrete values are a natural choice. Moreover, the parallel \ac{BF} algorithm is characterized by a very high algorithmic efficiency, which is an important requirement in code-based cryptography \cite{Baldi2018a, Misoczki2013}. Such an area of application is experiencing an increasing interest by the scientific community due to the standardization initiative of post-quantum cryptosystems started in 2016 by the US National Institute of Standards and Technology (NIST) \cite{NISTcall2016}. In this context, state-of-the-art schemes based on \ac{LDPC} and \ac{MDPC} codes such as LEDAcrypt \cite{LEDAcrypt} and BIKE \cite{BIKE} employ decoders such as \ac{BF} or some of its variants. This is all the more evident by considering that in these applications very large codes are usually required and the adoption of more complex decoding algorithms would yield unacceptable delays.

When \ac{LDPC} or \ac{MDPC} codes are used in code-based cryptosystems, the structure of their parity-check matrix is mainly dictated by security issues. This may yield unavoidable short cycles in the Tanner graph describing the code.
More precisely, in these systems the sparse parity-check matrix of an \ac{LDPC} or \ac{MDPC} code is used as a secret key and it usually has \ac{QC} structure. Starting from a code ensemble, according to the chosen \ac{QC} structure, the parity-check matrix of the code is randomly picked from the ensemble, thus often yielding a large number of cycles of length $6$ or even $4$. Accurate evaluation of the guaranteed error correction capability of codes with small girth has not been extensively investigated in previous literature. This is another relevant contribution of this paper, as we show that the new bounds are particularly tight if the girth of the considered codes is small.

We devote our attention to the first iteration of \ac{BF} decoding. For it, we provide an upper bound on the error rate of \ac{LDPC} and \ac{MDPC} codes which does not rely on any specific assumption. We note that some lower and upper bounds on the error rate under \ac{BF} decoding are also proposed in \cite{HuaXiao2006}, but their computation requires pre-processing of all possible initial error patterns with weight up to a certain value; thus, the approach becomes quickly unfeasible as the error probability of the channel decreases or error patterns with too large weight have to be considered. The same remark holds for the approaches proposed in \cite{Miladinovic2005,Chila2006,Xiao2007, Xiao2009}, which allow estimating the error rate of \ac{LDPC} codes under \ac{BF} decoding. Our approach instead is fully analytical, and does not require any preliminary simulation or assumption. To the best of our knowledge, this is the first time in which this problem is faced in exact analytical terms.

The paper is organized as follows. In Section \ref{sec:Notation} we introduce the notation used throughout the paper and recall some basic notions of \ac{LDPC} and \ac{MDPC} codes. In Section \ref{sec:Majority} we discuss the error correction capability of codes with small girth under \ac{BF} decoding. In Section \ref{sec:iters} we provide an upper bound on the error rate of \ac{LDPC} and \ac{MDPC} codes under \ac{BF} decoding. In Section \ref{sec:crypto} we present the results of numerical simulations and show an application of the derived bounds to code-based cryptography. Finally, we draw some conclusions in Section \ref{sec:conc}.

\section{Notation and definitions\label{sec:Notation}}
We use capital letters to denote sets, adopting caligraphic fonts for sets of vectors. The cardinality of a set $A$ (or $\mathcal{A}$) is denoted as $\card{A}$ (or $|\mathcal{A}|$). 
Given a set $A$, we use $a\gets A$ to express the fact that $a$ is randomly extracted, with uniform law, among all the elements of $A$, and the same notation is used for sets of vectors.

The binary Galois field is denoted as $\mathbb{F}_2$. We use small bold letters to denote vectors, and capital bold letters to denote matrices.
Given a matrix $\mathbf{H}$, its entry at position $(i,j)$ is denoted as $h_{i,j}$ and its $k$-th column is denoted as $\mathbf{h}_k$.  Given a vector $\mathbf{e}$, we refer to its $j$-th entry as $e_j$. 
Given a set $A$, we have $\mathbf{e}^{(A)}=\{e_i \st i\in A \}$. 
The AND, OR and ex-OR operations are denoted as $\wedge$, $\vee$ and $\oplus$, respectively.
The Hamming weight and the support of any vector $\be$ are referred to as $\weight{\mathbf{e}}$ and $S(\mathbf{e})$, respectively. The set of integers between $a$ and $b$, extremes included, is indicated as $[a,b]$. We denote the set of all binary vectors of length $n$ and Hamming weight $m$ as 
 $\mathcal{B}_m$.

\subsection{LDPC and MDPC codes}\label{subsec:LMDPC}

A binary \ac{LDPC} code is the null space of a binary parity-check matrix $\mathbf{H}$ containing a small number of ones compared to the total number of entries. Denoting the code block length as $n$ and the code dimension as $k$, $\bH$ has $r\geq n-k$ rows and $n$ columns and the design rate is $R=\frac{1}{2}$. 
The \textit{syndrome} of a binary vector $\mathbf{e}$ is defined as $\mathbf{s}=\mathbf{e}\mathbf{H}^{\top}$, where $^{\top}$ denotes transposition and the product is performed over $\mathbb{F}_2$. Any codeword belonging to the code defined by $\mathbf{H}$ has an all-zero syndrome. The $i$-th column and $j$-th row of $\bH$ have weight $v_i$ and $w_j$, respectively. The code is said to be $(v,w)$-\textit{regular} if each column of $\mathbf{H}$ contains exactly $v$ ones and each row contains exactly $w$ ones. Regular \ac{LDPC} codes are generally characterized by $w=O(\log{n})$, whereas regular \ac{MDPC} codes have $w=O(\sqrt{n})$. These two families of codes allow the same decoding principle, based on the sparsity of their parity-check matrices. 
Let us introduce two classes of \ac{QC} codes that will be considered throughout the paper (in particular, in Sections \ref{subsec:qccodes} and \ref{sec:crypto}). Codes in the first class are defined by parity-check matrices in the following form
\begin{equation}
\label{eq:double_circ}
\bH = \begin{bmatrix}\bH_0 & \bH_1\end{bmatrix},
\end{equation}
where each $\bH_i$, $i \in \{0, 1 \}$, is a circulant matrix of size $p$ and row/column weight $v$. The resulting codes are $(v,2v)$-regular, have block length $2p$ and design rate $R = \frac{1}{2}$.

 Codes in the second class, also named \emph{monomial codes} \cite{Fossorier2004}, are defined by parity-check matrices in the following form
\begin{equation}
\HH=\left[\begin{matrix}
\mathbf{I}^p(i_{0,0}) & \ldots & \mathbf{I}^p(i_{0,w-1})\\
\vdots & \ddots & \vdots\\
\mathbf{I}^p(i_{v-1,0}) & \ldots & \mathbf{I}^p(i_{v-1,w-1})\\
\end{matrix}\right],    
\label{eq:monocodes}
\end{equation}
where $\mathbf{I}^p(i)$ is the identity matrix of size $p$ whose columns have been cyclically shifted downwards by $i$ positions.

\begin{defn}\label{def:Gamma}
Given a matrix $\bH\in\mathbb{F}_2^{r\times n}$, the \emph{adjacency matrix} of $\bH$, denoted as $\mathbf{\Gamma}$, is the $n\times n$ matrix whose element in position $(i,j)$ is such that 
\[
\gamma_{i,j}=\begin{cases}
\card{S(\bh_i) \cap S(\bh_j)} \hspace{3mm} \text{\rm{if} $i\neq j$ }       \\
0 \hspace{29mm} \text{\rm{if} $i=j$}
\end{cases}.
\]

\end{defn}

The adjacency matrix is commonly employed in graph theory: given a multigraph with $n$ nodes, the adjacency matrix can be defined as the $n\times n$ matrix whose element in position $(i,j)$ is equal to the number of edges connecting nodes $i$ and $j$.
Obviously, starting from a parity-check matrix $\bH$, we can construct a graph\footnote{We remark that this graph, which is not bipartite, is different from the Tanner graph \cite{Tanner1981} of the code.} with $n$ nodes, such that the $i$-th and the $j$-th node are connected by $\left| S(\bh_i)\cap S(\bh_j)\right|$ edges.

\subsection{Bit flipping decoding}

Let us describe a general version of the parallel \ac{BF} algorithm, which performs a single iteration. 
Decoder inputs are a syndrome $\bs\in\mathbb{F}_2^r$ and a vector of integers $\mathbf{b}=[b_0,\cdots, b_{n-1}]$, such that $b_i \in [ 1 , v_i]$, $\forall i$. 
For each $i\in[0 , n-1]$,  the number of unsatisfied parity-check equations involving the $i$-th bit is computed; we denote such a number as $\sigma_i$.
The decoder considers as ``error affected'' all bits for which $\sigma_i\geq b_i$ and, thus, returns as output a vector $\be'$ with support $S(\be')=\{i\st \sigma_i\geq b_i\}$. So, $b_i$ has the meaning of a \textit{decision threshold} for the $i$-th bit. Clearly, decoding is successful if $\be'$ coincides with the actual error vector. An important special case considered next is that in which $b_i = b, \forall i$, which boils down to a majority-logic decoder when $b=\lfloor \frac{v}{2} \rfloor +1 $. The decoding procedure we consider is reported in Algorithm \ref{alg:BF}.
\begin{algorithm}[ht!]
\caption{{\fontfamily{cmss}\selectfont BFdecoder}}\label{alg:BF}
 \textbf{Input:}  $\bH\in\mathbb{F}_2^{r\times n}$, $\bs\in\mathbb{F}_2^r$, $i_{\texttt{max}}\in\mathbb{N}$, $\mathbf{b}=[b_0,\ldots,b_{n-1}]$, $b_i \in [ 1 , v_i]$, $\forall i$\\
\textbf{Output}: $\be' \in \mathbb{F}_2^n$\\
\begin{algorithmic}[1]
\State{$\be'\gets \mathbf{0}_{n}$}
\State{$F\gets \varnothing$}
\For{$i\gets 0\hspace{2mm}\textbf{to}\hspace{2mm}n-1$}
\State{$\sigma_i\gets 0$}
\For{$l\in S(\bh_i)$}
\State{$\sigma_i\gets \sigma_i+s_l$}
\EndFor
\If{$\sigma_i \geq b_i$}
\State{$F\gets F \cup i$}\Comment{Position $i$ is estimated as error affected}
\EndIf
\EndFor
\For{$i\in F$}
\State{$e'_i\gets e'_i \oplus 1 $}\Comment{Error estimation update}
\EndFor
\\\Return{$\{\be'\}$}
\end{algorithmic}
\end{algorithm}

\section{Guaranteed error correction capability of bit flipping \label{sec:Majority}}

Let us provide some preliminary definitions taken from \cite{Santini2019}, with some adaptations.

\begin{defn}
Given $\bH$, let us consider the rows of $\bH$ indexed by $S(\bh_i)$ and put them into a matrix $\bH^{(i)}$. Following \cite{Santini2019}, we define $\bH^{(i)}$ as the $i$-th \emph{partial parity-check matrix}. The $j$-th column of $\bH^{(i)}$ is denoted as $\bh_j^{(i)}$. We also define
\begin{equation*}
\delta^{(i)}(\bH^{(i)},z) = \max_{M, \hspace{1mm}\left|M\right|=z,\hspace{1mm}i\not\in M}{\left\{ \mathrm{wt}\Bigg( \bigoplus_{j \in M} \bh_j^{(i)} \Bigg) \right\}},
\end{equation*}
where $M$ is a set containing the indexes of $z$ columns of $\bH^{(i)}$, except for the $i$-th. We call the \emph{maximum column intersection of order $z$}, and denote as $\delta(\bH,z)$, the quantity defined as
\begin{equation*}
\delta(\bH,z) = \max_{ 0\leq i \leq n-1}{\left\{\delta^{(i)}(\bH^{(i)},z)\right\}}.
\end{equation*}
\end{defn}

When $z=1$, we call $\delta(\bH,1)$ the \emph{maximum column intersection} and, for simplicity, we denote it as $\delta$; it is easy to see that $\delta$ corresponds to the maximum number of set positions in which two columns of $\bH$ overlap.
We remark that, if the code has girth larger than $4$, then the supports of any two columns intersect in at most one position, thus we have $\delta=1$. 

The above notions can be easily related to the entries of the adjacency matrix.
For instance, the weight of the $j$-th column of the $i$-th partial parity-check matrix is equal to the $(i, j)$-th element of the matrix $\bGamma$, $\gamma_{i,j}$, and the maximum column intersection corresponds to the largest entry of $\bGamma$. For a code with girth larger than $4$, the entries of the adjacency matrix belong to $[0,1]$.

\begin{defn}
Given $\bH$ and the corresponding adjacency matrix $\bGamma$, we denote as $\btgamma{i}$ the vector formed by the elements of the $i$-th row of $\bGamma$, except for the $i$-th one. We define $\mu^{(i)}(z)$ as the sum of the $z$ largest entries of $\btgamma{i}$.
We then define the \emph{maximum column union of order $z$}, denoted as $\mu(\bH,z)$, the quantity 
\begin{equation}
\mu(\bH,z) = \max_{0\leq i \leq n-1} \Big\{\mu^{(i)}(z)\Big\}.
\label{eq:mu}
\end{equation}
\end{defn}

\subsection{Bounds on the error correction capability \label{subsec:boundserrcoca}}

The following theorem, from \cite{Tillich2018}, shows that the error correction capability of a code decoded with a majority-logic decoder is related to the maximum column intersection.

\begin{theorem}
\label{the:majority_Tillich}
{\rm \bf \cite{Tillich2018}}
Let us consider a code defined by a parity-check matrix for which every column has weight at least $v$ and whose maximum column intersection is $\delta$. Majority-logic decoding on this matrix allows the correction of all error vectors with weight $t \leq t_M$, where $t_M =\left\lfloor \frac{v}{2\delta}\right\rfloor$.
\end{theorem}

\begin{Cor}
\label{the:majority_Tillich_cor}
Let us consider a code with $g>4$ defined by a parity-check matrix for which every column has weight at least $v^*$. Majority-logic decoding on this matrix allows the correction of all error vectors with weight $t \leq t_M$, where $t_M =\left\lfloor \frac{v^*}{2}\right\rfloor$.
\end{Cor}
\begin{IEEEproof}
It is a straightforward consequence of the fact that, if $g>4$, the maximum column intersection is equal to $1$.
\end{IEEEproof}

As mentioned in the Introduction, these preliminary results are generalized in \cite{Santini2019}, where it is shown that the guaranteed error correction capability under \ac{BF} decoding can actually be expressed by taking into account the interplay of more than two columns, that is, assuming $z > 1$.

\begin{theorem}\label{the:majority_our}
{\rm \bf \cite{Santini2019}} Let us consider a code defined by a parity-check matrix $\bH$ in which every column has weight at least $v^*$.
Let $t$ be an integer such that
\begin{equation*}
v^* > \delta(\bH,t) + \delta(\bH,t-1).
\label{eq:condidelta}
\end{equation*}
Then a \ac{BF} decoder with  variable decoding thresholds \[b_i\in\left[\delta(\bH,t)+1, v_i-\delta(\bH,t-1) \right], \hspace{3mm} \forall i\in[0 , n-1], \hspace{3mm} v_i \geq v^*, \] (or fixed decoding threshold $b \in\left[\delta(\bH,t)+1, v^*-\delta(\bH,t-1) \right]$) corrects all the error vectors of weight $t$ in one iteration.
\end{theorem}

If we denote by $t_{M}$ the largest integer $t$ such  that  Theorem  \ref{the:majority_our} is satisfied, and assume that $\delta(\bH,i)\leq \delta(\bH,j), \footnote{This condition may be satisfied or not, depending on the structure of $\bH$.}\hspace{2mm}\forall i<j\leq t_M$, then Theorem  \ref{the:majority_our} allows correction of all the error vectors with weight smaller than or equal to $t_M$. 
Let us now specialize Theorem \ref{the:majority_our} to $(v,w)$-regular codes with girth $g>4$. When $g>4$, the weight of the columns of any partial parity-check matrix is either $0$ or $1$. In particular, any partial parity-check matrix contains one column with weight $v$, $(w-1)v$ columns with weight $1$ and $n-(w-1)v-1$ all-zero columns. As any partial parity-check matrix has $v$ rows, it follows that
\[
\delta(\bH,z) = z \quad \forall z \leq v,\\
\]
which is obtained by considering $z$ different columns. Then, according to Theorem \ref{the:majority_our}, we have that \[t_M=\max_t\left\{t \st v > t +t-1\right\}=\max_t\left\{t \st t\leq \left\lfloor \frac{v}{2}\right\rfloor \right\}=\left\lfloor \frac{v}{2}\right\rfloor,\]
with threshold $b=\left\lfloor\frac{v}{2}\right\rfloor+1$ if $v$ is even (corresponding to a majority-logic decoder), and $b\in[\left\lfloor\frac{v}{2}\right\rfloor+1, \left\lceil\frac{v}{2}\right\rceil+1]$ if $v$ is odd. 

In other words, when $g>4$, Theorem \ref{the:majority_Tillich} and Theorem \ref{the:majority_our} express the same error correction capability, with Theorem \ref{the:majority_our} giving an additional choice on the decision threshold when $v$ is odd. When $g=4$, instead, as proved in \cite{Santini2019}, the bound given in Theorem \ref{the:majority_our} is never smaller than that given in Theorem \ref{the:majority_Tillich}, which means that the new bound is tighter.

Theorem \ref{the:majority_our} guarantees correction of all error vectors up to a given weight $t_M$ only if $\delta(\bH,t)$ is a non-decreasing function for all $t\leq t_M$. This assumption is reasonable for sparse parity-check matrices, but it may be not verified for any choice of $\bH$; thus, we state the following Theorem \ref{the:majority_our_gamma}, based on the adjacency matrix $\bGamma$, which does not rely on any assumption. Theorem \ref{the:majority_our_gamma} provides an upper bound on the error correction capability that is smaller than or equal to the one given by Theorem \ref{the:majority_our}, but larger than or equal to the one given by Theorem \ref{the:majority_Tillich}.

\begin{theorem}\label{the:majority_our_gamma}
 Let us consider a code defined by a parity-check matrix $\bH$ in which every column has weight at least $v^*$.
Let $t$ be an integer smaller than or equal to $t_M$, where $t_M$ is the largest integer such that
\begin{equation}
v^* > \mu(\bH,t_M) + \mu(\bH,t_M-1).
\label{eq:condigamma}
\end{equation}
Then a \ac{BF} decoder with decoding thresholds 
\begin{equation}
b_i\in\left[\mu(\bH,t)+1, v_i-\mu(\bH,t-1) \right]
\label{eq:condigammab}
\end{equation}
corrects all the error vectors of weight smaller than or equal to $t$ in one iteration.
\end{theorem}
\begin{IEEEproof}
Let $\sigma_i$ denote the number of unsatisfied parity-check equations in which the $i$-th bit participates, and $v_i$ denote the weight of the $i$-th column in $\bH$. Let us denote by $\bf e$ the error vector and assume that $\weight{\be}=t$; if $e_i=1$, then we have
\begin{align}
\label{eq:counter_1}
\sigma_i^{(1)} \nonumber & = v_i - \mathrm{wt}\left(\bigoplus_{j \in S(\be)\setminus i}{\bh^{(i)}_j}\right)\\ 
& \geq v_i - \sum_{j\in S(\be)\setminus i}\gamma_{i,j}\\ \nonumber
& \geq v_i-\mu(\bH,t-1).
\end{align}
In the same way, when the $i$-th bit is error free, that is, $e_i=0$, we have
\begin{align}
\label{eq:counter_0}
\sigma_i^{(0)} \nonumber & = \mathrm{wt}\left(\bigoplus_{j \in S(\be)}{\bh^{(i)}_j}\right)\\
& \leq \sum_{j\in S (\be)}\gamma_{i,j}\\ \nonumber
& \leq \mu(\bH,t).
\end{align}

Clearly, one iteration of \ac{BF} decoding can correct any error vector $\be$ of weight $t$ if, $\forall i$, there exists a value of $b_i$ such that
\begin{equation}
\label{eq:nec_condition}
    \min_{\be} \{\sigma_i^{(1)}\} \geq b_i > \max_{\be} \{\sigma_j^{(0)}\}, \hspace{2mm}\forall i\in S(\be),\hspace{2mm}\forall j\not\in S(\be).
\end{equation}

Inserting \eqref{eq:counter_1} and \eqref{eq:counter_0} into \eqref{eq:nec_condition}, we obtain
\begin{equation}
v_i-\mu(\bH,t-1) \geq b_i > \mu(\bH,t),
\label{eq:threshold}
\end{equation}
which implies
\begin{equation}
v^*-\mu(\bH,t-1) > \mu(\bH,t).
\label{eq:condconta}
\end{equation} 

According to \eqref{eq:threshold}, any $b_i\in\left[ \mu(\bH,t)+1, v_i-\mu(\bH,t-1)\right]$ guarantees
that all bits such that $e_i=0$ are characterized by values of $\sigma_i^{(0)}$ that never exceed $b_i$ and, thus, are not flipped; oppositely, all bits such that $e_i=1$ are characterized by values of $\sigma_i^{(1)}$ larger than or equal to $b_i$, and thus are flipped.
\end{IEEEproof}

\subsection{Comparison with previous approaches}

In \cite{Chilappagari2008}, explicit formulas for bounds on the error correction capability are presented, thus we use them as a benchmark for our approach.
We remark that our bounds are referred to a single decoding iteration, whereas those in \cite{Chilappagari2008} are referred to an unspecified number of decoding iterations. Despite this, as shown in the following, for small values of $g$ our bounds are tighter than those in \cite{Chilappagari2008}. The latter are specified through the following theorem.

\begin{theorem}
{\rm \bf \cite{Chilappagari2008}} For a code defined by a parity-check matrix $\HH$ with girth $g$ in which every column has weight $v$, \ac{BF} decoding with decoding threshold $b=\left\lfloor\frac{v}{2}\right\rfloor+1$ allows correction of all error patterns of weight less than
\begin{equation}
\label{eq:Vasic}
\left\{ \begin{array}{cc}
\frac{1}{2} + \frac{v}{4} \sum_{i = 0}^{k - 1} \left( \frac{v - 2}{2} \right)^i & {\rm if}\; g = 4k + 2, \\
\sum_{i = 0}^{k - 1} \left( \frac{v - 2}{2} \right)^i & {\rm if}\; g = 4k. \\
\end{array}\right.
\end{equation}
\end{theorem}

For $g = 4$, $g = 6$ and $g=8$, the bounds on the error correction capability computed according to \eqref{eq:Vasic} are $0$, $\lceil \frac{v+2}{4} \rceil-1$ and $\lceil \frac{v}{2} \rceil-1$, respectively.
So, for $g=4$ \eqref{eq:Vasic} is useless. On the contrary, the error correction capability given by Theorem \ref{the:majority_our} is not null on condition that $\delta(\bH,0)+\delta(\bH,1)<v$, that is, being $\delta (\bH, 0) = 0$ by definition, if $\delta<v$. So, contrary to \eqref{eq:Vasic}, as long as $\bH$ does not contain repeated columns, Theorem \ref{the:majority_our} guarantees a significant error correction capability, just after one decoding iteration. Several examples are reported in \cite{Santini2019}, where it is shown that even the values resulting from Theorem \ref{the:majority_our_gamma} (that, we remind, are more conservative than those from Theorem \ref{the:majority_our}) are often significantly larger than those obtained from Theorem \ref{the:majority_Tillich}.

For $g = 6$, we have $\delta = 1$ and the error correction capability given by Theorem \ref{the:majority_our} coincides with that given by Theorem \ref{the:majority_our_gamma}, resulting in $t_M=\lfloor\frac{v}{2}\rfloor \geq \lceil \frac{v+2}{4} \rceil-1$.
Notice that the previous inequality, which compares the error correction capability given in Theorem \ref{the:majority_our_gamma} (left hand side) and that resulting from \eqref{eq:Vasic} (right hand side), holds  with the equality sign only for $v = 1$ and $v = 3$.  
To be more explicit, the gap between the correction capability foreseen by Theorem \ref{the:majority_our} and that obtained through \eqref{eq:Vasic} becomes higher and higher for increasing $v$, which is a significant issue in view of the application to code-based cryptography, where $v$ may assume relatively large values. 
Finally, for $g = 8$, Theorem \ref{the:majority_our} and Theorem \ref{the:majority_our_gamma} result in $\lfloor \frac{v}{2}\rfloor$, whereas \eqref{eq:Vasic} results in $\lceil \frac{v}{2} \rceil - 1$. So, since $\lfloor \frac{v}{2} \rfloor - \left(\lceil \frac{v}{2} \rceil - 1 \right) = 1 - v \; {\rm mod} \; 2$, the bounds are the same for odd values of $v$, whereas the bound we provide in Theorem \ref{the:majority_our} and Theorem \ref{the:majority_our_gamma} is larger by $1$ than that given in \eqref{eq:Vasic} for even values of $v$.

The comparison between the bounds we propose and those in \cite{Chilappagari2008} is summarized in Table \ref{tab:comparevane}, where by ``range of improvement'' we mean the values of $v$ for which our bound is strictly tighter than that in \cite{Chilappagari2008}.
\begin{table}[bt]
    \caption{Comparison of bounds on the error correction capability of \ac{LDPC} and \ac{MDPC} codes for different values of the girth.
    \label{tab:comparevane}}
    \centering
    \begin{tabular}{|c|c|c|c|}\hline
    $g$ & Bound on $t_M$ given by Theorem \ref{the:majority_our}  &  Eq. \eqref{eq:Vasic}  & Range of improvement \\\hline\hline
    4 & $\geq \lfloor\frac{v}{2\delta}\rfloor$  & 0 & $\forall v$ \\\hline
    6 & $\lfloor\frac{v}{2}\rfloor$ & $\lceil\frac{v+2}{4}\rceil-1$& $\forall v \neq 1,3$\\ \hline
    8 & $\lfloor\frac{v}{2}\rfloor$ & $\lceil\frac{v}{2}\rceil-1$ & $\forall v>2$, $v$ even \\ \hline
    $10$ & $\lfloor\frac{v}{2}\rfloor$ & $\lceil\frac{v^2+4}{8}\rceil-1$ & $v=2$ \\
    \hline
    \end{tabular}
\end{table}
The case of $g = 10$ has been also included in the table, for which the advantage of our approach is limited to the case of $v = 2$. The advantage disappears for $g > 10$ that, however, is not of interest in this paper.

So, based on the above considerations, we can conclude that the major impact of the present analysis and, similarly, of the analyses in \cite{Tillich2018, Santini2019}, occurs for codes with $g = 4$ and $g = 6$. 

\section{Analysis of the decoding failure probability for the first iteration of BF decoding\label{sec:iters}}

In this section we derive a conservative bound for the decoding failure probability, denoted as $P_f$,\footnote{Notice that the decoding failure probability coincides with the expected value of the \ac{FER}.} of the first and only iteration of a \ac{BF} decoder, with decoding thresholds $[b_0, b_1, \cdots, b_{n-1}]$, applied on a syndrome $\bs=\be\bH^{\top}$, where $\be\gets\Bt$.
Having a fixed number of errors ($t$) is a scenario of interest in code-based cryptography, in which encryption is performed by intentionally corrupting a codeword with a constant number of errors.
Nevertheless, once having characterized the decoder performance for a given number of errors, it is easy to extend such a characterization to channel models (like the \ac{BSC}) in which the statistic of the number of errors is known.
In fact, a \ac{BSC} with crossover probability $\rho$ can be straightforwardly studied by considering that the probability that the channel introduces exactly $t$ errors is equal to $\mathrm{Pr}\{\mathrm{wt}(\be)=t\}=\binom{n}{t}\rho^t (1-\rho)^{n-t}$. So, denoting the error vector after the first iteration as $\be'$, the decoding failure probability over the \ac{BSC} can be computed as
\begin{equation}
P_f=\sum_{l=0}^n{\mathrm{Pr}\left\{  \be'\neq \be \hspace{1mm}|\hspace{1mm} \mathrm{wt}(\mathbf{e})=l \right\} \mathrm{Pr}\left\{\mathrm{wt}(\mathbf{e}) = l\right\}},
\label{eq:FER1it}
\end{equation}
where $\mathrm{Pr}\left\{  \be'\neq \be \hspace{1mm}|\hspace{1mm} \mathrm{wt}(\mathbf{e})=l \right\}$ can be upper bounded through the method we describe next.
$\mathrm{Pr}\left\{\mathrm{wt}(\mathbf{e}) = l\right\}$, instead, defines the adopted channel model.
For the sake of conciseness, we only study the case in which
\[
\begin{cases}
\mathrm{Pr}\left\{\mathrm{wt}(\mathbf{e}) = l\right\} = 1, & l = t, \\
\mathrm{Pr}\left\{\mathrm{wt}(\mathbf{e}) = l\right\} = 0, & \forall l \neq t.
\end{cases}
\]
that models the application to code-based cryptography (where a fixed number $t$ of intentional errors is used for encryption).
However, our analysis can be easily extended to other channel models (like the \ac{BSC}) by changing the definition of $\mathrm{Pr}\left\{\mathrm{wt}(\mathbf{e}) = l\right\}$.

For $i\in[0 , n-1]$, we define $f_i$ as the binary variable obtained through the following rule
\begin{equation}
\label{eq:f_i}
    f_i = \begin{cases}0 & \text{if $[(\sigma_i<b_i)\wedge (e_i=0)] \vee [(\sigma_i\geq b_i)\wedge (e_i=1)]$},\\
    1 & \text{if $[(\sigma_i\geq b_i)\wedge (e_i=0)] \vee [(\sigma_i < b_i)\wedge (e_i=1)]$}.
    \end{cases}
\end{equation}
In other words, when $f_i=0$, the decoder takes a right decision on the $i$-th bit, i.e., it flips a bit affected by an error or it does not flip an error-free bit. Conversely, when $f_i=1$, the decoder takes a wrong decision on the $i$-th bit; a wrong decision can either be the flip of an error-free bit or the missing flip of a bit affected by an error.
The error patterns that cause a decoding error in the $i$-th position, that is, those for which $f_i=1$, are defined by the so-called \emph{error sets}, which we introduce below.

\begin{defn}\label{def:error_set}
Let $\bH\in\mathbb{F}_2^{r\times n}$ be the parity-check matrix of a code with block length $n$.
We consider the first and only iteration of a \ac{BF} decoder as in Algorithm \ref{alg:BF}, with decoding thresholds $[b_0,\cdots,b_{n-1}]$.
Let $f_i$ be the binary variable defined as in \eqref{eq:f_i}, for $i\in[0, n-1]$.
Then, for $z\in\{0 , 1\}$, we define the \emph{error set} for the $i$-th bit as follows 
\begin{equation*}
    \errset{z} = \left\{\left.\be\in\Bt\text{\hspace{1mm} s.t. \hspace{1mm}}f_i = 1 \right|e_i = z\right\}.
\end{equation*}
\end{defn}

As we show in the following section, the cardinality of each error set represents a fundamental quantity for assessing the error correction capability of the first iteration of a BF decoder as in Algorithm \ref{alg:BF}. Notice that the cardinality computation for each error set is strictly related to a subset sum problem, which in our case can be defined as follows: for a generic set, determine the number of subsets with given size having the property that the sum of their entries exceeds some target value.
The precise subset sum problem variant that we consider in this paper is formalized in the following definition.

\begin{defn}\label{def:subset_sum}
Let $\mathbf a\in\mathbb N^l$ be a length-$l$ vector.
For $m\leq l$, let $P_{l,m} = \{p_0,\cdots,p_{m-1}\}$ be a size-$m$ set of distinct integers in $[0, l-1]$ such that $p_0<p_1<\cdots p_{m-1}$.
Let $\mathcal{P}_{l,m}$ be the ensemble containing all such sets; clearly, $|\mathcal{P}_{l,m}| =\binom{l}{m}$. 
For $\alpha \in \mathbb{N}$, we define
\begin{equation*}
    \Nsub{\mathbf{a}}{m}{\alpha} = \left\{P_{l,m}\in\mathcal{P}_{l,m} \st      \sum_{i=0}^{m-1}{a_{p_i}} > \alpha\right\}.
\end{equation*}
\end{defn}

\subsection{Decoding failure probability analysis based on the error sets}

Let us introduce a property of the error sets that will then be used to derive the main result reported in Theorem \ref{the:general_theorem}.

\begin{Lem}
Let $\bH\in\mathbb{F}_2^{r\times n}$ be a parity-check matrix, and let $\errset{z}$, for $z\in\{0 , 1\}$, be the error set for the $i$-th bit.
We denote with $\btgamma{i}$ the vector formed by the entries of the $i$-th row of the adjacency matrix $\bGamma$, defined in Section \ref{sec:Notation}, except for the $i$-th one.
Then, we have
\begin{equation}
\label{eq:bound_1}
    \card{\errset{1}}\leq\card{\Nsub{\btgamma{i}}{t-1}{v_i-b_i}},\\
    \end{equation}
\begin{equation}
    \card{\errset{0}}\leq\card{\Nsub{\btgamma{i}}{t}{b_i-1}}.
    \label{eq:bound_0}
\end{equation}
\end{Lem}
\begin{IEEEproof}
We focus on the $i$-th bit, characterized by a certain value of $\sigma_i$ and flipping threshold $b_i$, and derive the conditions upon which the decoder takes a wrong decision (i.e., $f_i = 1$).
We first consider the case of $e_i=1$: a wrong decision is taken if the decoder does not flip the bit, i.e., if $\sigma_i < b_i$.
From \eqref{eq:counter_1}, we know that the value of $\sigma_i$ is not lower than the difference between the weight of the $i$-th column (that is, $v_i$) and the sum of the values $\gamma_{i,j}$ indexed by $S(\be)$, except the $i$-th index (that is, $\sum_{j\in S(\be)\setminus{i}}{\gamma_{i,j}}$).
If such a difference is not lower than $b_i$, then $\sigma_i\geq b_i$ and the decoder flips the $i$-th bit.
On the other hand, if
$v_i-\sum_{j\in S(\be)\setminus{i}}{\gamma_{i,j}} < b_i$, $\sigma_i$ might be lower than $b_i$ and the decoder might not flip the $i$-th bit.
Hence, a necessary (but not sufficient) condition to have a wrong decision on the $i$-th bit is \begin{equation}
\label{eq:proof_bound}
\sum_{j\in S(\be)\setminus{i}}{\gamma_{i,j}} > v_i- b_i.    
\end{equation}
Because of the above reasoning, $\errset{1}$ is a subset of the error vectors satisfying \eqref{eq:proof_bound}.
The set $S(\mathbf e)\setminus{i}$ in \eqref{eq:proof_bound} corresponds to a subset of $[0 , i-1]\cup [i+1, n-1]$, of size $t-1$; furthermore, the values $\gamma_{i,j}$ that are possibly selected by $S(\mathbf e)\setminus i$ are entries of
$\tilde{\boldsymbol{\gamma}}^{(i)}=[\gamma_{i,0}, \cdots, \gamma_{i,i-1},\gamma_{i,i+1}, \cdots, \gamma_{i,n-1}]$, which has length $n-1$.
Let $P_{n-1,t-1}$ be a subset of $[0 , n-1]$ such that the sum of the entries in $\tilde{\boldsymbol{\gamma}}^{(i)}$ indexed by $P_{n-1,t-1}$ is larger than $v_i-b_i$.
According to Definition \ref{def:error_set}, the number of such sets corresponds to the cardinality of $\mathcal N^{\tilde{\boldsymbol{\gamma}}^{(i)}}_{t-1,v_i-b_1}$.
Furthermore, to each one of these subsets, we can associate an error vector satisfying \eqref{eq:proof_bound}, with support
$$\left\{j \in P_{n-1,t-1}\left|j<i\right.\right\}\cup i \cup \left\{j+1 \in P_{n-1,t-1}\left|j>i\right.\right\}.$$
Thus, we obtain
\begin{align*}
\label{eq:fi_1_1}
\card{\errset{1}}\nonumber & \leq \card{\left\{  \be\in\Bt\text{\hspace{1mm}s.t.\hspace{1mm}}  (e_i = 1)\wedge \left(\sum_{j\in S(\be)\setminus{i}}{\gamma_{i,j}} > v_i-b_i\right)\right\}}\\\nonumber
& = \card{ \Nsub{\btgamma{i}}{t-1}{v_i-b_i}}.
\end{align*}
Similarly, for the case of $e_i=0$, we can derive from \eqref{eq:counter_0} that a necessary but not sufficient condition for $f_i=1$ is $b_i\leq \sigma_i\leq \sum_{j\in S(e)}{\gamma_{i,j}}$. Similarly to the case of $e_1=1$, we have
\begin{align*}
\card{\errset{0}} & \leq \card{\left\{  \be\in\Bt\text{\hspace{2mm}s.t.\hspace{2mm}} (e_i = 0) \wedge  \left(\sum_{j\in S(e)}{\gamma_{i,j}} > b_i-1\right)\right\}}\\
& = \card{\Nsub{\btgamma{i}}{t}{b_i-1}}.
\end{align*}

\end{IEEEproof}

Based on these relationships, we can now prove the following main theorem.

\begin{theorem}\label{the:general_theorem}
Let $\bH\in\mathbb{F}_2^{r\times n}$ be a parity-check matrix. 
Let $\be\in\Bt$, and $\bs = \be\bH^T$ be the corresponding syndrome.
We consider a single BF iteration applied on $\bs$, with decoding threshold for the $i$-th bit denoted as $b_i$. 
     Let  $\btgamma{i}$ denote the vector formed by the elements in the $i$-th row of $\bGamma$, except for the $i$-th one. 
The probability that the decoder fails to decode, starting from $\bs$, is upper bounded as follows
\begin{equation}
\label{eq:Th4}
    P_f \leq \min\left\{1 ;  \frac{\sum_{i=0}^{n-1} \left( |\Nsub{\btgamma{i}}{t-1}{v_i-b_i}|+|\Nsub{\btgamma{i}}{t}{b_i-1}| \right)} {\binom{n}{t}}\right\}.
\end{equation}
\end{theorem}
\begin{IEEEproof}
Let us start from an arbitrary position $i\in[0 , n-1]$.  
Let $\mathcal{E}_{i,t,b_i}$ be the set of error vectors of weight $t$ such that, when the decoding threshold for the $i$-th bit is $b_i$, the decoder decision results in $f_i = 1$ (i.e., the decoder flips the bit if $e_i=0$ or does not flip the bit if $e_i=1$).
Clearly $\mathcal{E}_{i,t,b_i}=\errset{0}\cup\errset{1}$.
Moreover, the sets $\errset{1}$ and $\errset{0}$ are disjoint, since the vectors in $\mathbf e\in\errset{1}$ are such that $e_i=1$ and those in $\errset{0}$ are such that $e_i=0$. Taking into account \eqref{eq:bound_1} and \eqref{eq:bound_0}, we obtain
\begin{align}
\label{eq:bound_E}
\card{\mathcal{E}_{i,t,b_i}} \nonumber & = \card{\mathcal{E}^0_{i,t,b_i}} + \card{\mathcal{E}^1_{i,t,b_i}}\\
& \leq \card{\Nsub{\btgamma{i}}{t-1}{v_i-b_i}} + \card{\Nsub{\btgamma{i}}{t}{b_i-1}}.
\end{align}
For all values $j\in[0 , n-1]$ such that $\mathcal{E}_{j,t,b_j}$ contains $\mathbf e$, we have $f_j = 1$, i.e., a wrong decoder decision is taken on the $j$-th bit.
Then, the probability that decoding fails can be upper bounded by means of the following chain of inequalities
\begin{align}
\label{eq:bound_derivation}
    P_f \nonumber & = \frac{\left|\bigcup_{i=0}^{n-1}\mathcal{E}_{i,t,b_i}\right|}{|\Bt|}\\\nonumber
    & \leq  \frac{\sum_{i=0}^{n-1}\left|\mathcal{E}_{i,t,b_i}\right|}{|\Bt|}\\
    & \leq \frac{\sum_{i=0}^{n-1}\left( \card{\Nsub{\btgamma{i}}{t-1}{v_i-b_i}} + \card{\Nsub{\btgamma{i}}{t}{b_i}} \right)}{\card{\Bt}}.
\end{align}
The thesis of the theorem is finally proved by considering that $\card{\Bt}=\binom{n}{t}$ and that, by definition, $P_f\leq 1$ (while the bound in \eqref{eq:bound_derivation} is not guaranteed to be smaller than or equal to $1$).
\end{IEEEproof}

In order to compute the bound given in the theorem above, we need to solve instances of the subset sum problem according to Definition \ref{def:subset_sum}.
Clearly, the naive approach of testing all possible subsets of vectors $\tilde{\boldsymbol{\gamma}}^{(i)}$ is computationally unfeasible.
Fortunately, in our case of interest, the problem can be eased by considering that, due to the sparsity of the parity-check matrix, $\tilde{\boldsymbol{\gamma}}^{(i)}$ is likely to contain a large number of very small entries (the majority of which being actually null). 
This peculiarity of sparsity makes the problem efficiently solvable; a low complexity approach to perform this computation is described in Appendix A.

The expression of $P_f$ derived above is coherent with the results given in Section \ref{subsec:boundserrcoca} and, in particular, in Theorem \ref{the:majority_our_gamma}. Indeed, the following corollary holds.
\begin{Cor}
Let us suppose that $t\leq t_M$, where $t_M$ is the largest integer such that \eqref{eq:condigamma} holds. If the decoding threshold is chosen as follows
\begin{equation}
    b_i\in\left[\mu(\bH,t)+1, v_i-\mu(\bH,t-1) \right], \hspace{3mm} \forall i,
    \label{eq:sogliagamma}
\end{equation}
 then $\card{\Nsub{\btgamma{i}}{t-1}{v_i-b_i}} = \card{\Nsub{\btgamma{i}}{t}{b_i}} = 0$, $\forall i$ and, consequently, $P_f=0$.
\end{Cor} 
\begin{IEEEproof}
By definition, 
\begin{align*}
\card{\Nsub{\btgamma{i}}{t-1}{v_i-b_i}} &=\card{ \left\{P_{n,t-1}\in\mathcal{P}_{n,t-1} \st      \sum_{i=0}^{t-2}{\btgamma{i}_{p_i}} > v_i-b_i\right\}}\\
&=\card{ \left\{P_{n,t-1}\in\mathcal{P}_{n,t-1} \st b_i> v_i-\sum_{i=0}^{t-2}{\btgamma{i}_{p_i}}\right\}}.\\
\end{align*}
However, it follows from the definition of $\mu(\bH,t-1)$ and from \eqref{eq:sogliagamma} that \[b_i\leq v_i- \mu(\bH,t-1)\leq v_i- \sum_{i=0}^{t-2}{\btgamma{i}_{p_i}}\] for any choice of the indexes $p_i$ and, thus, $\card{\Nsub{\btgamma{i}}{t-1}{v_i-b_i}}=0$.
Similarly, we have
\begin{align*}
\card{\Nsub{\btgamma{i}}{t}{b_i-1}} &=\card{ \left\{P_{n,t}\in\mathcal{P}_{n,t} \st      \sum_{i=0}^{t-1}{\btgamma{i}_{p_i}} > b_i-1\right\}}\\
&=\card{ \left\{P_{n,t-1}\in\mathcal{P}_{n,t-1} \st b_i<\sum_{i=0}^{t-1}{\btgamma{i}_{p_i}}+1 \right\}}.\\
\end{align*}
It also follows from \eqref{eq:sogliagamma} that
\[
b_i\geq \mu(\bH,t)+1\geq \sum_{i=0}^{t-1}{\btgamma{i}_{p_i}}+1
\]
for any choice of the indexes $p_i$, and thus $\card{\Nsub{\btgamma{i}}{t}{b_i-1}}=0$. Finally, the fact that $P_f=0$ is a straightforward consequence of \eqref{eq:Th4}.
\end{IEEEproof}

In the particular case of regular codes, which implies to have equal decoding threshold values, noted as $b$, assuming $v$ is odd and $b = \bmaj$, the bound on $P_f$ provided by Theorem \ref{the:general_theorem} can be rewritten as
\begin{equation}
\label{eq:Th4bis}
    P_f \leq \min\left\{1 ;  \frac{\sum_{i=0}^{n-1}\card{\Nsub{\boldsymbol{\gamma}^{(i)}}{t}{\frac{v-1}{2}}}}{\binom{n}{t}}\right\}.
\end{equation}
The proof is reported in Appendix B. 

Equation \eqref{eq:Th4bis} can be used for any regular code with $g \ge 4$. 
For regular codes with $g \geq 6$, however, \eqref{eq:Th4bis} can be further elaborated as discussed next.

\subsection{Regular codes with girth larger than $4$}

When $g \geq 6$, we have
\begin{equation}
    \gamma_{i,j}\in\{0 , 1\}, \hspace{2mm}\forall i,j.
\end{equation}
In particular, for $(v,w)$-regular codes, each row and each column of $\bGamma$ contain exactly $v(w-1)$ non-zero entries. The following lemma holds.

\begin{Lem}
\label{lem:teta}
Let $\ba\in\mathbb{F}_2^l$ be a vector of weight $m$; then, we have $\card{\Nsub{\ba}{x}{\alpha}} = \theta(l,x,m,\alpha)$, with
\begin{equation}
    \theta(l,x,m,\alpha) = \begin{cases} 0 & \text{if $\alpha \geq m$ \hspace{2mm} or\hspace{2mm} $x \leq \alpha$ }\\
    \sum_{j = \alpha + 1}^{\min\{m , x\}}\binom{m}{j}\binom{l-m}{x-j}& \text{otherwise}
    \end{cases}.
\label{eq:theta}
\end{equation}
\end{Lem}

The following Theorem \ref{the:girth6} specializes Theorem \ref{the:general_theorem} to the case of a regular code with girth larger than $4$, and reformulates \eqref{eq:Th4bis} for such a case.

\begin{theorem}\label{the:girth6}
Let $\bH\in\mathbb{F}_2^{r  \times n}$ be the parity-check matrix of a $(v,w)$-regular code with girth $g\geq 6$. 
Let $\be\in\Bt$, and $\bs = \be\bH^\top$.
We consider a single  iteration of \ac{BF} decoding applied to $\bs$, with a unique decoding threshold $b$.
If $v$ is odd and $b = \bmaj$, we have
\begin{equation}
    \begin{cases}
    P_f = 0 & \text{if $t \leq \frac{v-1}{2}$} \\
   P_f \leq \min\left\{1 ;  \frac{n\theta(n,t,v(w-1), \frac{v-1}{2})}{\binom{n}{t}} \right\} & \text{otherwise}\\
    \end{cases},
    \label{eq:bound6nosets}
\end{equation} where, using \eqref{eq:theta}, \[\theta\Big(n,t,v(w-1),\frac{v-1}{2}\Big)= \sum_{j = \frac{v+1}{2}}^{\min\{v(w-1) , t\}}\binom{v(w-1)}{j}\binom{n-v(w-1)}{t-j}.\]
\end{theorem}
\begin{IEEEproof}
The proof is quite similar to that of Theorem \ref{the:general_theorem} and its specialization to the case of regular codes (reported in Appendix B),
by taking into account Lemma \ref{lem:teta}.
\end{IEEEproof}

\subsection{A special class of \ac{QC} codes} \label{subsec:qccodes}

In this section we consider \ac{QC} codes with parity-check matrix as in \eqref{eq:double_circ}, which is interesting for cryptographic applications, as will be discussed in Section \ref{sec:crypto}. By considering the \ac{QC} nature of these codes, described by parity-check matrices made of circulant blocks, the bounds introduced in the previous sections can be further specialized.
It can be easily verified that, for these codes, the matrix $\bGamma$ is \ac{QC} as well; this property can be exploited to further speed-up the computation of the error sets required to calculate the bounds.

The following well-known result holds.

\begin{Lem}
Any circulant matrix with weight larger than $2$ has girth $g\leq 6$.
\label{lem:6cycles}
\end{Lem}
\begin{IEEEproof}
The proof is omitted for brevity. See \cite[Lemma 4.2]{Baldi2014a}.
\end{IEEEproof}

It follows from Lemma \ref{lem:6cycles} that a parity-check matrix as in \eqref{eq:double_circ} cannot have girth larger than $6$. 

In this case, the matrix $\bGamma$ can be written as
\begin{equation}
\label{eq:Gamma}
    \bGamma = \begin{bmatrix}
    \bGamma_{0,0} & \bGamma_{0,1}\\
    \bGamma_{1,0} & \bGamma_{1,1}
    \end{bmatrix},
\end{equation}
where each $\bGamma_{i,j}$ is a $p\times p$ matrix; in particular, $\bGamma$ is symmetric, and this means that $\bGamma_{0,0}$ and $\bGamma_{1,1}$ are symmetric as well, while $\bGamma_{0,1}^\top=\bGamma_{1,0}$. 
Moreover, each block $\bGamma_{i,j}$ is circulant.
In particular, let $\bgamma{i}$ be the $i$-th row of $\bGamma$; then, all rows $\bgamma{j}$ such that $\left\lfloor i/p \right\rfloor = \left\lfloor j/p \right\rfloor$ are identical up to a quasi-cyclic shift; this means that
\begin{equation}
    \label{eq:QC_gamma}
    \card{\mathcal{E}^{z}_{i,t,b}} = \card{\mathcal{E}^{z}_{j,t,b}}, \hspace{2mm}\forall b,t,\hspace{2mm}\forall i,j\text{\hspace{1mm}s.t.\hspace{1mm}}\left\lfloor i/p \right\rfloor = \left\lfloor j/p \right\rfloor,
\end{equation}
with $z\in\{0 , 1\}$.
Then, from Theorem \ref{the:general_theorem} we obtain
\begin{equation}
\label{eq:QC_DFR}
    P_f \leq \min\left\{1 ;  p\hspace{1mm}  \frac{\mathcal{N}_{\mathrm{tot}}}{\binom{n}{t}}\right\},
\end{equation}
with
\[\mathcal{N}_{\mathrm{tot}}=\card{\Nsub{\btgamma{0}}{t-1}{v-b}}+\card{\Nsub{\btgamma{0}}{t}{b-1}} + \card{\Nsub{\btgamma{p}}{t-1}{v-b}}+\card{\Nsub{\btgamma{p}}{t}{b-1}}.\]

\section{Application to cryptography}
\label{sec:crypto}

In this section we assess the accuracy of our bound through numerical simulations. Then, we make some considerations on the connections of the proposed bound with the security levels of code-based cryptosystems.

\subsection{Numerical simulations}
\label{subsec:numres}

There is a recent trend in post-quantum cryptography regarding the use of \ac{QC-LDPC} and \ac{QC-MDPC} codes \cite{Baldi2012,Misoczki2013,Baldi2018a} defined in Section \ref{subsec:qccodes}, since they enable the design of McEliece cryptosystem variants with very small public keys. We remark that, in code-based cryptography, a decoding failure yields a decryption failure; thus, the \ac{FER} coincides with the so-called \ac{DFR}.

Let us first consider some codes defined by parity-check matrices as in \eqref{eq:double_circ}. 
In order to show the tightness of the provided bounds, let us consider different choices of code parameters.
First, we analyze some specifically designed codes, whose column weight is chosen in such a way as to approach or reach the expected guaranteed error correction capability through Monte Carlo simulations.
Then, we also consider codes that have actually been proposed for cryptographic applications, whose column weight must be sufficiently large to withstand key recovery attacks~\cite[Section 5.2]{Misoczki2013}.

\begin{figure}
    \centering
    \includegraphics[keepaspectratio, width=9cm]{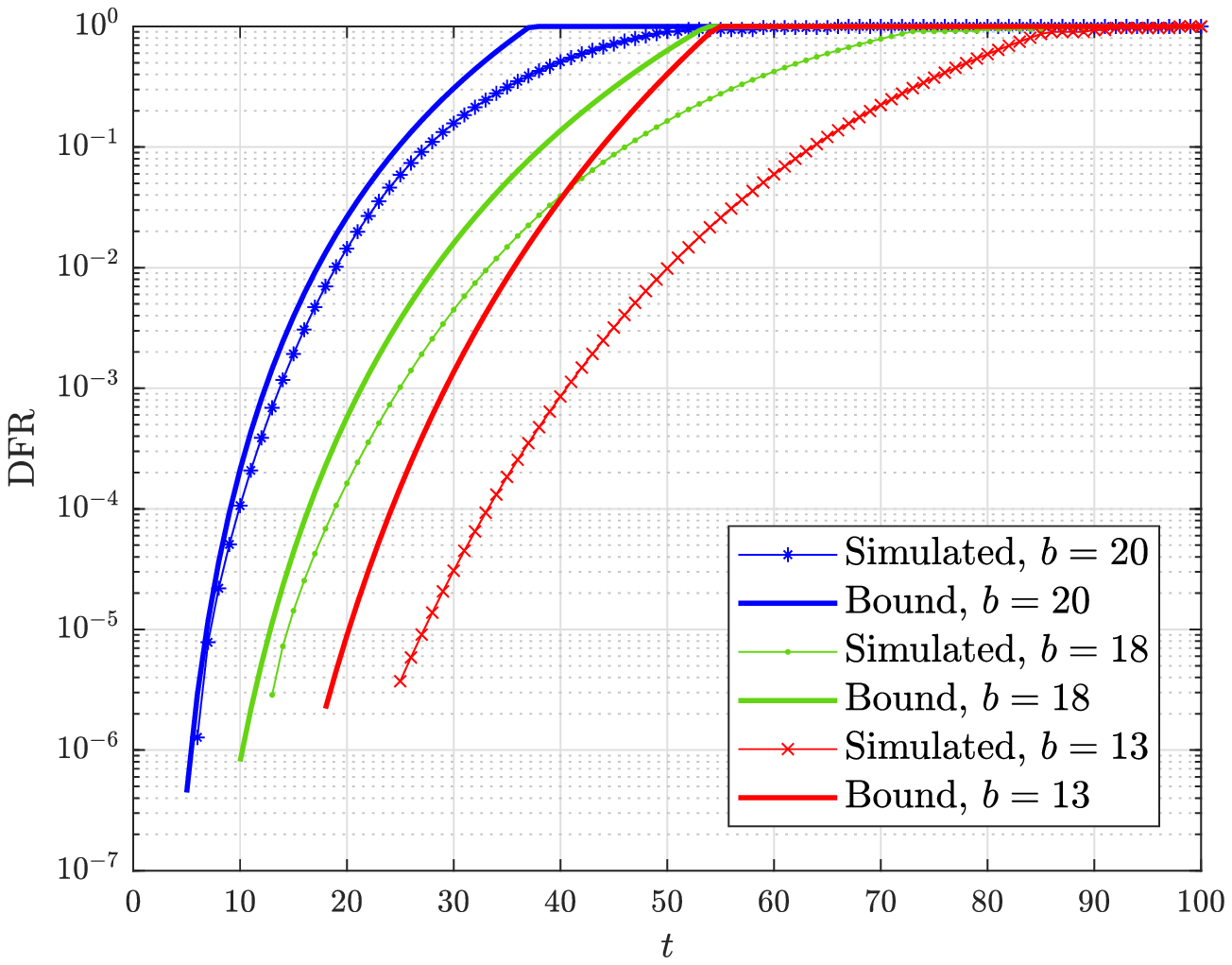}
    \caption{Comparison of the \ac{DFR} resulting from Monte Carlo simulations with our bound for a $(25,50)$-regular code with block length $n=19\,702$, $R=\frac{1}{2}$, $p=9\,851$, $v=25$, $g=4$, and different threshold values.} 
    \label{fig:p5483}
\end{figure}
\begin{figure}
    \centering
    \includegraphics[keepaspectratio, width=9cm]{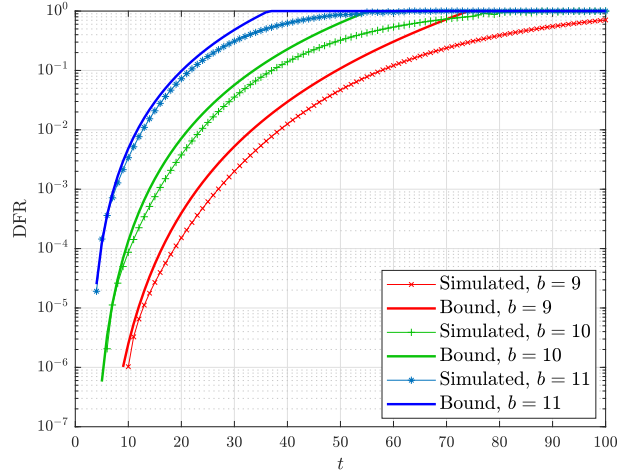}
    \caption{Comparison of the \ac{DFR} resulting from Monte Carlo simulations with our bound, for a $(13,26)$-regular code with block length $n=17\,558$, $R=\frac{1}{2}$, $p=8\,779$, $v=13$, $g=6$, and different threshold values.}
    \label{fig:p8779}
\end{figure}

 In order to assess the behaviour of codes with similar parameters and different girth, let us consider a first code, $\mathcal{C}_0$, with length $n=19\,702$, design rate $R=\frac{1}{2}$, $p=9\,851$, $v=25$, $g=4$ and a second code, $\mathcal{C}_1$, with $n=17\,558$, design rate $R=\frac{1}{2}$, $p=8\,779$, $v=13$ and girth $g = 6$. A compact representation of their parity-check matrices is available in Appendix C.
We assess the \ac{DFR} achieved by a single-iteration \ac{BF} decoder with different threshold values through Monte Carlo simulations; for each value of $t$, the \ac{DFR} has been estimated through the observation of $100$ wrong decoding instances. 
The comparison of the simulation results with our bounds is shown in Figs. \ref{fig:p5483} and \ref{fig:p8779}, respectively. 
From the figures we observe that for both codes the bound becomes tighter and tighter for decreasing values of $t$.

\begin{figure}
    \centering
    \includegraphics[keepaspectratio, width=9cm]{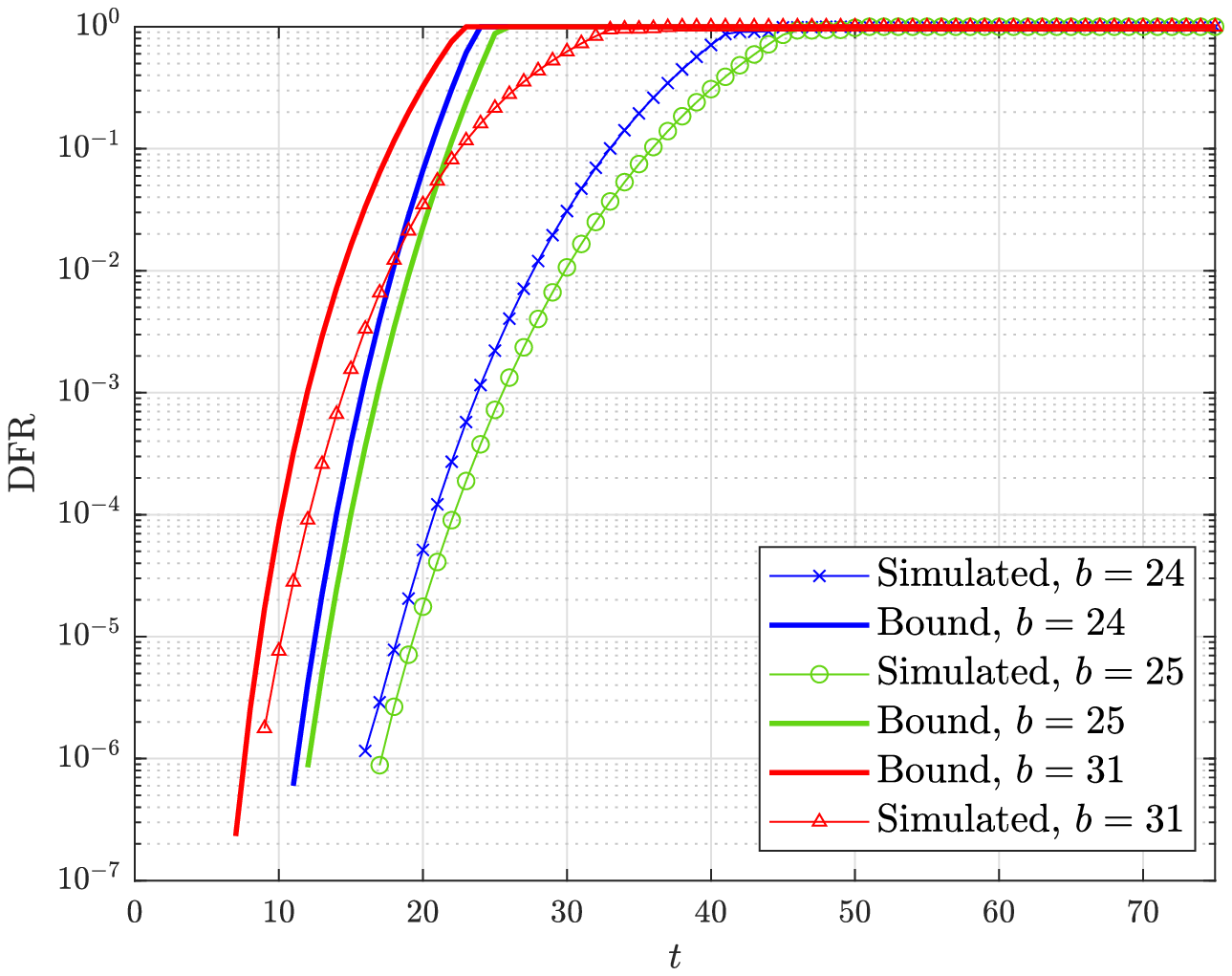}
    \caption{Comparison of the \ac{DFR} resulting from Monte Carlo simulations with our bound for a $(45,90)$-regular code with block length $n=9\,602$, $R=\frac{1}{2}$, $p=4\,801$, $g=4$, and different threshold values.} 
    \label{fig:ccode}
\end{figure}

Let us now consider a $(45,90)$-regular code, $\mathcal{C}_2$, with block length $n=9\,602$, circulant block size $p=4\,801$, design rate $R=\frac{1}{2}$ and girth $g=4$. These parameters are suitable for cryptographic applications \cite{Misoczki2013}. A compact representation of its parity-check matrices is available in Appendix C. Also in this case, its error rate performance is compared to the bound, considering different thresholds. The results are shown in Fig. \ref{fig:ccode}. We notice that, also in this case, the bound becomes tighter and tighter for decreasing values of $t$. 

\begin{figure}
    \centering
    \includegraphics[keepaspectratio, width=9cm]{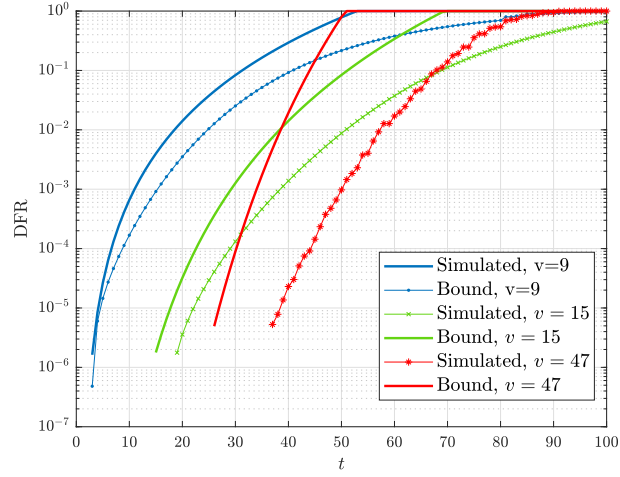}
    \caption{Comparison of the \ac{DFR} resulting from Monte Carlo simulations with our bound, for $(v,2v)$-regular codes with block length $n=23\,434$, $R=\frac{1}{2}$, $p=11\,717$, $v\in\{9,15,47\}$, $g=4$, and $b=\lfloor\frac{v}{2}\rfloor+1$.}
    \label{fig:p11717diffv}
\end{figure}

In order to assess the effect of the parity-check matrix column weight, let us consider three $(v,2v)$-regular codes, $\mathcal{C}_3$, $\mathcal{C}_4$ and $\mathcal{C}_5$, defined by parity-check matrices as in \eqref{eq:double_circ}, with the same block length, $n=23\,434$, circulant block size $p=11\,717$, and design rate $R=\frac{1}{2}$, but different values of the column weight: $v = 9$, $v = 15$ and $v = 47$ for $\mathcal{C}_3$, $\mathcal{C}_4$ and $\mathcal{C}_5$, respectively. A compact representation of their parity-check matrices is available in Appendix C.  The decoding threshold is chosen as $b = \lfloor \frac{v}{2} \rfloor +1$. The simulation results are shown in Fig. \ref{fig:p11717diffv}. Also in these cases, the bound becomes tighter and tighter for decreasing values of $t$. We also remark that the bound is tight for both \ac{LDPC} and \ac{MDPC} codes; in fact, $\mathcal{C}_3$ and $\mathcal{C}_4$ are \ac{LDPC} codes, whereas $\mathcal{C}_5$ is an \ac{MDPC} code.

\begin{figure}
    \centering
    \includegraphics[keepaspectratio, width=9cm]{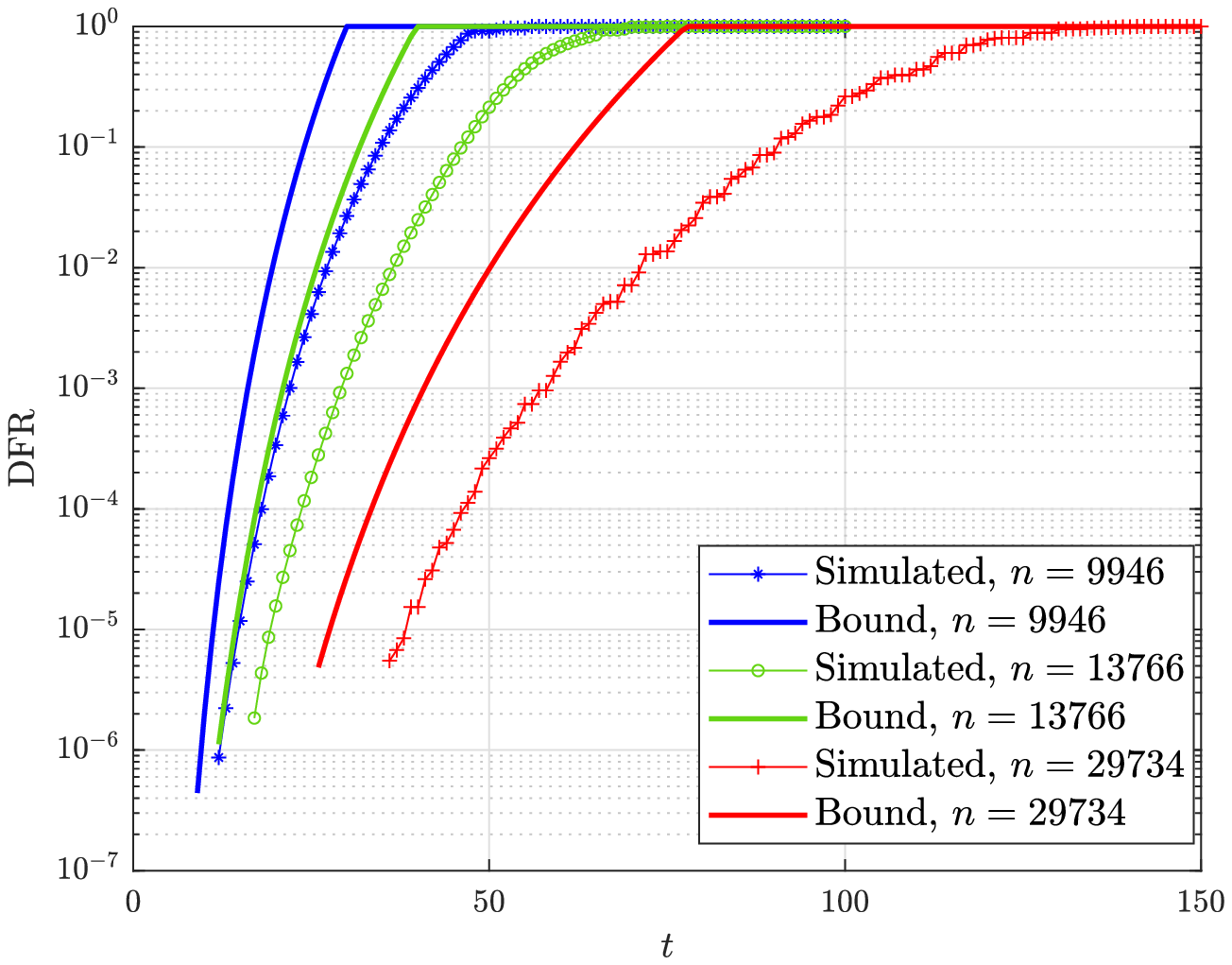}
    \caption{Comparison of the \ac{DFR} resulting from Monte Carlo simulations with our bound, for $(25,50)$-regular codes with block length $n\in\{9\,946,13\,766,29\,734\}$, $R=\frac{1}{2}$, $p=\frac{n}{2}$, $g=4$, and $b=13$.}
    \label{fig:v25diffn}
\end{figure}

In order to assess the effect of the block  length, let us fix the parity-check matrix row and column weight and consider three $(25,50)$-regular codes, $\mathcal{C}_6$, $\mathcal{C}_7$ and $\mathcal{C}_8$, defined by parity-check matrices as in \eqref{eq:double_circ}, with block length $n = 9\,946$, $n = 13\,766$ and $n = 29\,734$, respectively. A compact representation of their parity-check matrices is available in Appendix C. Also in this case, the threshold is $b = \lfloor \frac{v}{2} \rfloor +1$. A comparison of their \ac{DFR} with the proposed bound is shown in Fig. \ref{fig:v25diffn}. In all these cases, the bound becomes tighter and tighter for decreasing values of $t$, as in the previously considered cases. 


\begin{figure}
    \centering
    \includegraphics[keepaspectratio, width=9cm]{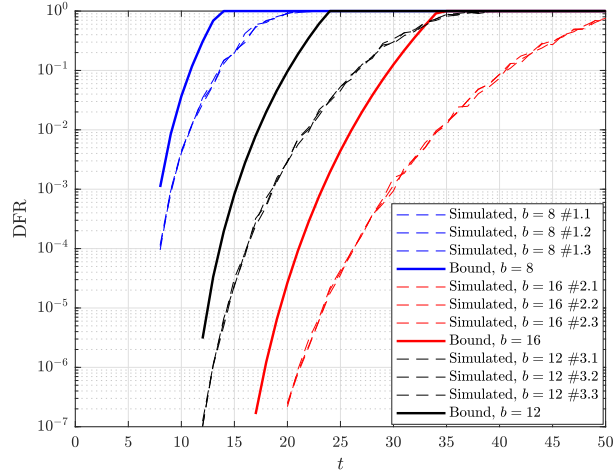}
    \caption{Comparison of the \ac{DFR} resulting from Monte Carlo simulations with our bound, for monomial codes described by the parameters in Table \ref{table:Tabparam}.}
    \label{fig:mono}
\end{figure}

Finally, let us consider a different family of codes, that is, monomial codes defined in Section \ref{subsec:LMDPC}. It is shown in \cite{Santini2018ISIT} that, for a proper choice of the shifts and of the code parameters, monomial codes can be used in code-based cryptosystems. Thus, we consider \ac{QC-LDPC} codes of this type designed through the technique  suggested in \cite[Section IV-C]{Santini2018ISIT} with some modifications, in such a way as to obtain codes with variable rate and row/column weight. These codes have girth $6$ and design rate $R=1-\frac{v}{w}$, and we assess their error rate performance considering $b=\lceil\frac{v}{2} \rceil$, as imposed by Theorem \ref{the:girth6}.
 In particular, let us consider three parameter sets, described in Table \ref{table:Tabparam}, and for each parameter set, i.e., for each code ensemble, we randomly generate three monomial codes and compare their error rate performance with the bound given by \eqref{eq:bound6nosets}. Results are shown in Fig. \ref{fig:mono}. We observe that there is no appreciable difference between the performance of codes in the same ensemble. We also observe that the bound is tight for monomial codes as well.

\begin{table}[!t]
\renewcommand{\arraystretch}{1}
\caption{Parameters of the considered monomial codes.}
\label{table:Tabparam}
\centering
\begin{tabular}{|c|c|c|c|c|c|c|c|c|}
\hline
Parameter Set & $n$ & $r$ & $v$ & $w$& $p$ & $g$ & Design rate\\ \hline\hline
 \# 1 & $4\,171$ & $1\,455$ & $15$ & $43$ & $97$ & $6$ & $0.65$\\ \hline
 \# 2 & $8\,517$ & $5\,177$ & $31$ & $51$ & $167$ & $6$ & $0.39$\\ \hline
  \# 3 & $3\,937$ & $2\,921$ & $23$ & $31$ & $127$ & $6$ & $0.26$\\ \hline
\end{tabular}
\end{table}

\subsection{Design of codes with given \ac{DFR}}
\label{subsec:crypto}

When codes as in \eqref{eq:double_circ} are used in code-based cryptosystems that support key reuse, the required values of $P_f$ are much smaller than those reported in the figures of Section \ref{subsec:numres}, and are impossible to assess through Monte Carlo simulations.
In particular, in order to avoid key recovery attacks based on decryption failures, such as those in~\cite{Guo2016},\cite{ Santini2019a}, also called \emph{reaction attacks}, a cryptosystem designed for a $2^{\lambda}$ security level (expressed as number of binary operations) must have $\mathrm{DFR}<2^{-\lambda}$~\cite{Hofheinz2017} with values of $\lambda$ not smaller than $80$.
A negligible decoding failure probability is also required to achieve the desirable security condition known as \ac{IND-CCA} \cite{Hofheinz2017}.

This makes the derived bounds particularly useful in this case.
In fact, by assuming the \ac{QC} code structure specified in Section \ref{subsec:qccodes}, we can use \eqref{eq:QC_DFR} to design code parameters able to achieve the desired small values of $P_f$ without requiring any simulation.
To show an example, let us consider the case of a security level of $2^{80}$ binary operations, for which \ac{QC-MDPC} codes with $v \geq 45$ and $t\geq 84$ are needed \cite{Misoczki2013}.
The matrices proposed in \cite{Misoczki2013} have $p=4801$, which however leads to a decoding failure probability too large to resist reaction attacks and to achieve \ac{IND-CCA}. A decoding failure probability lower than $2^{-80}$ is instead required for such a purpose.

Indeed, the bound given in \eqref{eq:QC_DFR} allows achieving such a requirement through a classic rejection sampling approach: for each randomly generated parity-check matrix in the form \eqref{eq:double_circ}, the bound \eqref{eq:QC_DFR} is computed and the matrix discarded if such a value is above the target $P_f$. The procedure is repeated until a matrix with the desired property is obtained.
In order to verify the feasibility of such an approach, we consider different parameter sets and, for each set, we generate $1\,000$ parity-check matrices at random and compute the bound on $P_f$ given by \eqref{eq:QC_DFR}. The choice of $b$ is optimized by choosing its value for which the bound takes its minimum.
 
The results of this experiment are reported in Table \ref{tab:tabpars}. We notice that, for all tested parameter sets, a significant percentage of matrices satisfies the constraint $P_f< 2^{-80}$. 
This fact guarantees that 
the time required to generate a valid matrix is limited.
In other words, it is not difficult to find a matrix for which we can be sure that the desired security level is reached.

We point out that, despite the codes obtained through the above approach are significantly larger than those originally proposed, they still lead to public key sizes that are smaller than those of other competing cryptosystems, while achieving \ac{IND-CCA}.
For instance, considering binary Goppa codes as in the original McEliece cryptosystem, the public key size equals $460\,647$ bits \cite{Bernstein2008} for $80$ bits security, while the parameters we have found lead to a reduction in the public key size by a factor ranging between $1.64$ and $3.57$.  
Additionally, the parameter sets we propose represent a concrete worst case estimate of the key size increase which is needed in order to ensure \ac{IND-CCA}.
Indeed, we obviously expect that if more than one decoding iteration is performed, the minimum value of $p$ which is necessary to fulfill $P_f<2^{-80}$ decreases, thus further reducing the key size and allowing more significant improvements with respect to other cryptosystems. However, extending the bound to the case of multiple iterations goes beyond the scope of this paper and is left for future works.
 
\begin{table}[ht]
    \caption{Number of selected keys 
    for different parameter sets. 
    \label{tab:tabpars}}
    \centering
    \begin{tabular}{|c|c|c|}\hline
    $p$ & $v$ &  Keys achieving $P_f < 2^{-80}$ \\\hline\hline
    $279\,991$ & 45 & 158 out of 1\,000 \\\hline
    194\,989 & 65 & 990 out of 1\,000 \\\hline
    160\,499 & 75 & 792 out of 1\,000 \\\hline
    149\,993 & 85 & 971 out of 1\,000 \\\hline
    138\,389 & 95 & 847 out of 1\,000 \\\hline
    130\,043 & 105 & 226 out of 1\,000 \\\hline
    \end{tabular}

\end{table}

\section{Conclusion \label{sec:conc}}

We have studied the error correction capability of \ac{LDPC} and \ac{MDPC} codes under \ac{BF} iterative decoding, with the aim of finding theoretical models for its characterization without resorting to computation-intensive simulations.

Under the simplifying setting of a single-iteration \ac{BF} decoder, we have shown that a per-code upper bound on the error rate can indeed be found.
Such a bound provides an important tool in those contexts where very small error rates have to be guaranteed for each specific code.

One of these scenarios is that of code-based cryptography, and we have shown how our bound can be succesfully applied to such a context, allowing the design of cryptosystems based on \ac{QC-LDPC} and \ac{QC-MDPC} codes able to achieve strong security notions while keeping the size of the public keys smaller than that of classic systems employing algebraic codes and bounded-distance decoders.



\section*{Appendix A}
\label{sec:Appendix_A}

In this appendix we describe an efficient way to compute the cardinalities of the sets introduced in Definition \ref{def:subset_sum}.
To this end, we first formalize the problem and then describe a method that, for the cases we are interested in, significantly improves upon the naive exhaustive search approach.

\begin{prob}\label{def:prob_sum}
Let $\ba\in\mathbb{N}^l$ be a length-$l$ vector of non negative integers, and let $B\subseteq [0 , l-1]$ be a set  of size $m\leq l$.
Given $\alpha\in\mathbb{N}$, $\alpha>0$, compute
\begin{equation*}
N_B = \left|\left\{B\subseteq [0 ,l-1],\hspace{2mm} |B|=m\hspace{2mm}\text{s.t.}\hspace{2mm} \sum_{i\in B}a_i > \alpha \right\}\right|.
\end{equation*}
\end{prob}
It is clear that an exhaustive search would require to generate all subsets of size $m$: thus, the corresponding complexity will be equal to $\binom{l}{m}$.
As we show with combinatorial arguments, a simple algorithm can be devised, with a complexity that may be significantly lower.

In particular, we obtain the number of sets that are complementary to those defined in Problem \ref{def:prob_sum}, that is,
\begin{equation*}
\bar{N}_B = \left|\left\{B\subseteq [0 ,l-1],\hspace{2mm} |B|=m\hspace{2mm}\text{s.t.}\hspace{2mm} \sum_{i\in B}a_{i} \leq \alpha \right\}\right|,
\end{equation*}
from which the value of $N_B$ can be straightforwardly obtained as
\begin{equation}
    N_B = \binom{l}{m} - \bar{N}_B.
\end{equation}
For a set $B$, we denote with $\mathbf{a}^{(B)}$ the vector formed by the entries of $\mathbf{a}$ that are indexed by $B$; we define $\bar{N}_B^{(j)}$ as the number of subsets $B$ for which the corresponding sub-vector $\mathbf{a}^{(B)}$ contains $m$ elements, $j$ of which are distinct,  whose sum is smaller than or equal to $\alpha$.
We have 
\begin{equation}
\label{eq:compl_sums_barNb_j}
    \bar{N}_B = \sum_{j=1}^{m}\bar{N}_B^{(j)}.
    \end{equation}
The values of $\bar{N}^{(j)}_B$ can be easily obtained, as we show next.

First of all, let $\omega$ be the number of distinct values in $\mathbf{a}$, with
$Y = \{y_0,y_1,\cdots,y_{\omega-1}\}$ being the set of such values in ascending order.
In the same way, we define $\lambda_u = \left|\left\{i \hspace{2mm}\text{s.t.}\hspace{2mm}a_i = y_u\right\}\right|$.
As we show below, the computation of $\bar{N}_B$ depends only on these quantities.

Let $Y_B$ be the set of distinct values that are contained in $\mathbf{a}^{(B)}$. 
When $j = 1$, we easily have
\begin{equation}
\label{eq:Nb_1}
    \bar{N}_B^{(1)} = \sum_{\begin{smallmatrix} 0 \leq i \leq \omega-1 \hspace{1mm} : \hspace{1mm} y_i\leq \left\lfloor \frac{\alpha}{m} \right\rfloor\\
   \end{smallmatrix}}^{}\binom{\lambda_i}{m},
\end{equation}
where, as usual, $\binom{\lambda_i}{m}=0$ if $m>\lambda_i$.
When $j>1$, some further considerations must be taken into account.
For a set $B$, let $y_{i_0}, y_{i_1},\cdots,y_{i_{j-1}}$ be the distinct values assumed by the entries of $\mathbf{a}^{(B)}$, and denote the corresponding multiplicities as $m_0,m_1,\cdots,m_{j-1}$.
If $B \in \bar{N}_B^{(j)}$, we must have 
\begin{equation}
\label{eq:subset_sum}
    \sum_{u=0}^{j-1}m_{u}y_{i_u}\leq\alpha.
\end{equation}
We clearly have $m=\sum_{u=0}^{j-1}m_u$, from which we obtain $m_0 = m - \sum_{u=1}^{j-1}m_u$; then,  \eqref{eq:subset_sum} can be rewritten as
\begin{equation}
    my_{i_0} + \sum_{u=1}^{j-1}{m_u(y_{i_u}-y_{i_0})}\leq \alpha.
\end{equation}
It is obvious that
\begin{equation}
my_{i_0} + \sum_{u=1}^{j-1}{m_u(y_{i_u}-y_{i_0})}\geq my_{i_0} + \sum_{u=1}^{j-1}{(y_{i_u}-y_{i_0})}.
\end{equation}
The above condition can be turned into the following criterion: a set $B$ associated to the values $y_{i_0}, y_{i_1},\cdots,y_{i_{j-1}}$ of $\mathbf{a}^{(B)}$, whose sum is smaller than or equal to $\alpha$, exists if and only if
\begin{equation}
\label{eq:condition_A}
\sum_{u=1}^{j-1}{(y_{i_u}-y_{i_0})}\leq \alpha - my_{i_0}. 
\end{equation}

Let us now fix an index $q\in [1 , j-2]$, and suppose that we are looking at all sets $B$ such that $\mathbf{a}^{(B)}$ contains the values $y_{i_0},\cdots,y_{i_{q-1}}$ with respective multiplicities $m_1, m_2, \cdots, m_{q-1}$.
Then, imposing the constraint and summing over all subsets, we obtain
\begin{align*}
    \alpha \geq my_{i_0} + & \nonumber \sum_{u=1}^{q-1}m_u(y_{i_u}-y_{i_0}) + m_q(y_{i_q}-y_{i_0}) + \sum_{z=q+1}^{j-1}m_z(y_{i_z}-y_{i_0})\\\nonumber
    & \geq my_{i_0} + \sum_{u=1}^{q-1}m_u(y_{i_u}-y_{i_0}) + m_q(y_{i_q}-y_{i_0}) + \sum_{z = q+1}^{j-1}(y_{i_z}-y_{i_0}).
\end{align*}
Then, the maximum value for $m_q$ is obtained as
\begin{equation}
\label{eq:condition_B}
m_q^{(\texttt{max})} = \min\left\{\lambda_q , \left\lfloor \frac{\alpha-my_{i_0} - \sum_{u=1}^{q-1}m_u(y_{i_u}-y_{i_0}) - \sum_{z=q+1}^{j-1}(y_{i_z}-y_{i_0})}{y_{i_q}-y_{i_0}}\right\rfloor\right\}.
\end{equation}
Finally, $\bar{N}_B^{(j)}$ can be computed as
\begin{equation}
\label{eq:barNb_j}
    \bar{N}_B^{(j)} = \sum_{i_0 = 0}^{\omega-j}\sum_{i_1 = i_0+1}^{\omega-j+1} \cdots \sum_{i_{j-1} = i_{j-2}+1}^{\omega-1}d(i_0,\cdots,i_{j-1}),
\end{equation}
where 

\begin{equation}
\label{eq:d_coeff}
    d(i_0,\cdots,i_{j-1}) = \begin{cases}
        0 \hspace{24mm} \text{if $\sum_{u=1}^{j-1}{(y_{i_u}-y_{i_0})} > \alpha - my_{i_0} $}\\
        \sum_{m_1 = 1}^{m_1^{(\texttt{max})}}\cdots \sum_{m_{j-1} =1}^{m_{j-1}^{(\texttt{max})}}\binom{\lambda_{i_0}}{m-\sum_{i=1}^{j-1}m_i}\prod_{u=1}^{j-1}\binom{\lambda_{i_u}}{m_u} & \text{otherwise}
\end{cases}.
\end{equation}

We point out that when $\ba$ contains a small number of distinct elements (i.e., $\omega \ll l$) this approach becomes significantly faster than the exhaustive search on all subsets.
Indeed, first of all we clearly have $\bar{N}_B^{(j)} = 0$ when $j > \omega$; moreover, the number of configurations tested by using \eqref{eq:d_coeff} is surely smaller than $m^{j-1}$.
Then, for a specific value of $j$, the computation of $\bar{N}_B^{(j)}$ requires to test no more than $m^{j-1}\binom{\omega}{j}$ configurations.  
Thus, we can roughly upper bound the total number of configurations that are considered as
\begin{equation}
\sum_{j=1}^{\omega}{m^{j-1}\binom{\omega}{j}}\leq \sum_{j=1}^{\omega}m^{j-1}\left( \frac{\omega e}{j}\right)^j\leq \omega m^{\omega-1} e^{\omega},
\end{equation}
where $e$ is the basis of the natural logarithmic.
It can be verified that, when $m,w\ll l$, the above upper bound is significantly smaller than $\binom{l}{m}$.

\section*{Appendix B}
\label{sec:Appendix_B}

In this appendix we consider the case of regular codes, for which the decoding threshold values can be assumed constant and equal to $b$, and we demonstrate that when $v$ is odd and $b=\bmaj$, the bound \eqref{eq:Th4} can be reformulated as in \eqref{eq:Th4bis}.

Let $\bH$ be the parity-check matrix of a $(v,w)$-regular code with block length $n$ and odd $v$. Let us denote as $\bgamma{i}$ the $i$-th row of the adjacency matrix $\boldsymbol{\Gamma}$.
Moreover, let $\be\in \Bt$, and $\bs = \be\bH^\top$.
We consider a single  iteration of \ac{BF} decoding applied to $\bs$, with a unique decoding threshold $\bmaj$.

In order to determine a bound for $P_f$ in these conditions, we can basically repeat the steps in the proof of Theorem \ref{the:general_theorem}. In this case, however,
\eqref{eq:bound_E} can be specialized as
follows
\begin{align}
\card{\mathcal{E}_{i,t,\bmaj}}\nonumber &  = \card{\mathcal{E}^{1}_{i,t-1,v-\bmaj}\cup \mathcal{E}^{0}_{i,t,\bmaj-1}}\\\nonumber
& = \card{\mathcal{E}^{1}_{i,t-1,\frac{v-1}{2}}\cup \mathcal{E}^{0}_{i,t,\frac{v-1}{2}}}\\
& \leq \card{\Nsub{\tilde{\boldsymbol{\gamma}}^{(i)}}{t-1}{\frac{v-1}{2}}} + \card{\Nsub{\tilde{\boldsymbol{\gamma}}^{(i)}}{t}{\frac{v-1}{2}}},
\end{align}
where we have exploited the fact that, since $v$ is odd, we have $\bmaj = \frac{v+1}{2}$.
Now, if we consider $\bgamma{i}$ and a set $S\in\Nsub{\bgamma{i}}{t}{\frac{v-1}{2}}$, we have only two possibilities:
\begin{enumerate}
    \item If $i\in S$, since $\gamma_{i,i} = 0$, we have
$\sum_{j\in S\setminus i}\gamma_{i,j} > \frac{v-1}{2}$, from which $\{S\setminus i\} \in \Nsub{\btgamma{i}}{t-1}{\frac{v-1}{2}}$.
\item If $i\not\in S$, we have
$\sum_{j\in S}\gamma_{i,j} > \frac{v-1}{2}$, from which $S\in \Nsub{\btgamma{i}}{t}{\frac{v-1}{2}}$.
\end{enumerate}
Then, we can state
\begin{equation}
    \card{\Nsub{\btgamma{i}}{t-1}{\frac{v-1}{2}}} + \card{\Nsub{\btgamma{i}}{t}{\frac{v-1}{2}}}=\card{\Nsub{\bgamma{i}}{t}{\frac{v-1}{2}}}.
\end{equation}
By replacing this equality in \eqref{eq:Th4}, the simpler \eqref{eq:Th4bis} is eventually obtained.

\section*{Appendix C}
\label{sec:Appendix_C}

In this appendix we give the parity-check matrices used in the Monte Carlo simulations. All the considered matrices are in form \eqref{eq:double_circ} and $\bH_0$ and $\bH_1$ are circulant matrices.  The support of their first columns, which is $S(\bh_0^{(0)})$ and $S(\bh_0^{(1)})$, but is denoted for simplicity as $S_0$ and $S_1$, respectively, compactly describes the whole parity-check matrix.

The parity-check matrix of $\mathcal{C}_0$ is represented by
\begin{multline*}
 S_0=\small\left[\begin{array}{ccccccccccccc}
 16   &      364         &572&        1166 &       1726        &2231&        2518 &       2555 &       2565&       3334 &       3806&  3818 &       4126 \end{array} \right.\\
\small\left.\begin{array}{ccccccccccccc}      4590    &    4852 &       5425 &       5502       & 5536 &       5576      &  5880     &   7923 &       8296  & 8788        &9035  &      9179
\end{array}
\right],
\end{multline*}
\begin{multline*}
S_1=\small\left[\begin{array}{ccccccccccccc}
   246  &       406 &       1732    &    1855&        1871 &       2254     &   2297        &2320&        2474 &       3333 &       3513
&   4042      \end{array} \right.\\
\small\left.\begin{array}{ccccccccccccc}     4511  &      5260   &     6037 &       6673  &      6716    &    7334  &      7766     &   7940   &    8036  &      8136 & 8802  &      8881  &      9384
\end{array}
\right].   
\end{multline*}
The parity-check matrix of the code $\mathcal{C}_1$ is represented by
\begin{multline*}
 S_0=\left[\small\begin{array}{ccccccccccccccccc}
  934     &   1750   &     3485 &       4040     &   4117  &      4639       & 4838 & 4879    &    5874      &  5886    &    6041 &  6874    &    7425
\end{array}
\right],
\end{multline*}
\begin{multline*}
S_1=\left[\small\begin{array}{ccccccccccccc}
   2043   &     2184   &     2619      &  2715   &     3190  &      3359   &     4163   &   4327      &  4705   &     5188 &       5335   &      7629    &    7879
\end{array}
\right].
\end{multline*}

The parity-check matrix of $\mathcal{C}_2$ is represented by
\begin{multline*}
S_0=\left[\small\begin{array}{ccccccccccccc}
       168 & 229 & 309 & 405 & 464 & 507 & 668 & 888 & 893 & 908 & 984 & 1015\end{array} \right.\\ \left.\small\begin{array}{ccccccccccc}1143&1178&1299&1311&1368&1380&1433&1478&1675&1728&1800\end{array} \right.\\ \left.\small\begin{array}{ccccccccccc}1936&2069&2084&2215&2530&2632&2842&3090&3103&3282&3332\end{array} \right.\\ \left.\small\begin{array}{ccccccccccc}3532&3595&3657&3882&3919&3929&4077&4138&4160&4654&4698
\end{array}\right],    
\end{multline*}
\begin{multline*}
S_1=\left[\small\begin{array}{cccccccccccc}
        263      &   271    &     277    &     369   &      381 &        641   &      689&        754       &  792 &        935 &1153       & 1415 
        \end{array} \right.\\ \left.\small\begin{array}{cccccccccccc}
        1551 &       1727       & 1732    &    1743&        1988      &  2065&        2099    &    2102 &2139      &  2159      &  2205    
        \end{array} \right.\\ \left.\small\begin{array}{cccccccccccc}
        2249    &    2443  &      2566    &    2586      &  2737 &       2932    &    3041 & 3140  &      3337&       3504    &    3613   
        \end{array} \right.\\ \left.\small\begin{array}{ccccccccccc}3632  &      3946 &        3953 &        4047  &      4097    &    4218 & 4233     &   4315  &      4329       & 4486   &     4506
\end{array}\right].
\end{multline*}

The parity-check matrix of $\mathcal{C}_3$ is represented by
\[
S_0=\small\left[\begin{array}{ccccccccc}
       864  &      3551      &  4164    &    5538       & 8013  &      8487     &   8846&        8986   &    10925
\end{array}\right],
\]
\[
S_1=\small\left[\begin{array}{ccccccccc}
        2256    &    6346       & 6495   &     6959     &   7551      &  8409    &    8725      & 10317     &  11554
\end{array}\right].
\]

The parity-check matrix of $\mathcal{C}_4$ is represented by
\begin{multline*}
S_0=\left[\small\begin{array}{ccccccccccccccc}
       1106   &     1985        &2497&        3036       & 3394      &  5118  &      5136   &     5276 \end{array} \right.\\
\left.\small\begin{array}{ccccccccccccc}        6506    &    6523& 7450&        8338  &      8472  &      9662  &     11434
\end{array}\right],
\end{multline*}
\begin{multline*}
S_1=\left[\small\begin{array}{ccccccccccccccc}
471      &   974    &    1775        &5048  &      5595     &   5617  &      6805     &   8861   \end{array} \right.\\
\left.\small\begin{array}{ccccccccccccc}    8894    &    9009    &    9158  &      9416      & 11071   &    11379   &    11404
\end{array}\right].
\end{multline*}

The parity-check matrix of $\mathcal{C}_5$ is represented by
\begin{multline*}
S_0=\left[\small\begin{array}{ccccccccccccccc}
      242       &  432 &        447 &        784    &    1040    &    1669&        1786    &    2430    &    2496     &   & 2643 & 2682  &     3161 &        3173   \end{array} \right.\\
\left.\small\begin{array}{ccccccccccccc}     3952    &    4461 &       5319     &   5336   &     5369  &      5423&        5678   &     5768
&
5891   &     6906   &     6943       & 7207&        7535   \end{array} \right.\\
\left.\small\begin{array}{ccccccccccccc}       7740 &       7743    &    8435     &   8496        &8608 &8765& 8824 &      9251&        9463  &      9635    &    9637 &       9659     &   9685  \end{array} \right.\\
\left.\small\begin{array}{ccccccccccccc}       9969  &      9971     &  10052
&
10284     &  10397  &     10525    &   10821&       11367
\end{array}\right],
\end{multline*}
\begin{multline*}
S_1=\small\left[\begin{array}{ccccccccccccccc}
     144      &   284   &     722    &     724    &     821    &    1403   &     1465    &    1546       & 2028   &     2277
&  
2569     &      2916  &      3108        \end{array} \right.\\
\small\left.\begin{array}{ccccccccccccc}    3286    &    3400 &       3460 &       3759 &       3844        &3983
& 
4252    &    4600   &     4631  &      5289 &      5323     &   5587    &    6004    \end{array} \right.\\
\small\left.\begin{array}{ccccccccccccc}         6403&   7380 &       7427
&
 7826   &     7899      &  7998 &       8106      &      8960 &       9004     &   9196    &   9348    &    9508 &
9803   \end{array} \right.\\
\small\left.\begin{array}{ccccccccccccc} 10058& 10497  &     10671  &     10751 &      10865     &  11092    &  11362    &   11394
\end{array}\right].
\end{multline*}

The parity-check matrix of $\mathcal{C}_6$ is represented by
\begin{multline*}
S_0=\small\left[\begin{array}{ccccccccccccccc}
 516        &   739    &       988   &       1332     &     1408     &     1503     &     1668      &    1671      &    1743    &      1983
 &  
      2042        &  2110     &     2466          \end{array} \right.\\
\small\left.\begin{array}{ccccccccccccc}    2583        &  2661        &  2808     &     2863    &      2918     &     2976   &       3388
&
   3551    &    3828   &     4337   &     4533   &     4741
\end{array}\right],
\end{multline*}
\begin{multline*}
S_1=\left[\small\begin{array}{ccccccccccccccc}
 132     &    448    &     502    &       769     &      868     &     1063     &     1436      &    1457     &     1511  &       1676
  & 2023     &     2422       &   2469        \end{array} \right.\\
\small\left.\begin{array}{ccccccccccccc}      2613   &       2620       &   3197    &      3499       &   3754  &        4020    &      4054 &
 4211      &    4286     &     4528       &   4599     &     4930
\end{array}\right].
\end{multline*}

The parity-check matrix of $\mathcal{C}_7$ is represented by
\begin{multline*}
S_0=\small\left[\begin{array}{ccccccccccccccc}
  709     &    792  &       854     &    907    &    1548&        1608  &      2062      &  2152  &      2158      &  2359
  &  
 2625     &   2981      &  3372      \end{array} \right.\\
\small\left.\begin{array}{ccccccccccccc}    3572&        3664&        3716   &     3726   &     4283   &     5311   &     5551
&6014    &    6432     &   6569   &     6595     &   6636
\end{array}\right],
\end{multline*}
\begin{multline*}
S_1=\small\left[\begin{array}{ccccccccccccccc}
 824   &      934   &     1220   &     1570   &     2129  &      2244    &    2526    &    2629    &    3533    &    3557
&
   3708    &    3833    &    3862     \end{array} \right.\\
\small\left.\begin{array}{ccccccccccccc}    4147     &   4252  &      4556  &      4636     &   4662    &    5254   &     5286
&
 5375    &    5691  &      5738 &       6347    &    6785
\end{array}\right].
\end{multline*}

The parity-check matrix of $\mathcal{C}_8$ is represented by
\begin{multline*}
S_0=\small\left[\begin{array}{ccccccccccccccc}
1383   &     1783  &      1940  &      2117     &   2834 &        3216    &    3347   &     4168   &     4267      &  6118
   &
  7683  &      8431     &   9114     \end{array} \right.\\
\small\left.\begin{array}{ccccccccccccc}     9191 &       9562&       10170  &     10515     &  10874    &   11604 &      12110
&
 13137  &     13202  &     13508  &     14658 &      14687
\end{array}\right],
\end{multline*}
\begin{multline*}
S_1=\small\left[\begin{array}{ccccccccccccccc}
  189    &     272      &   753     &    938    &    1372 &        1940    &    1984      &  2524     &   3072   &     4414
&4637  &      4807     &   4971     \end{array} \right.\\
\small\left.\begin{array}{ccccccccccccc}     6029  &      6360&        6931   &     6970      &  7653    &    8817 &       9193& 11761    &   11981 &      12242     &  12549  &     13846
\end{array}\right].
\end{multline*}

\bibliographystyle{IEEEtran}
\bibliography{Archive}

\end{document}